# Photomolecular Effect: Visible Light Interaction with Air-Water Interface

Guangxin Lv[1], Yaodong Tu[1], James H. Zhang, Gang Chen[*]

**Affiliations:**

Department of Mechanical Engineering, Massachusetts Institute of Technology, Cambridge, MA 02139.

*Corresponding author. Email: gchen2@mit.edu.

[1]These authors contributed equally to the work.

**Abstract:** Although water is almost transparent to visible light, we demonstrate that the air-water interface interacts strongly with visible light via what we hypothesize as the photomolecular effect. In this effect, transverse-magnetic polarized photons cleave off water clusters from the air-water interface. We use over 10 different experiments to demonstrate the existence of this effect and its dependence on the wavelength, incident angle and polarization of visible light. We further demonstrate that visible light heats up thin fogs, suggesting that this process can impact weather, climate, and the earth's water cycle. Our study suggests that the photomolecular effect should happen widely in nature, from clouds to fogs, ocean to soil surfaces, and plant transpiration, and can also lead to new applications in energy and clear water.

**One-Sentence Summary:** Photons in the visible spectrum where bulk water normally does not absorb can cleave off large water clusters from the air-water interface.

**Main Text:**

Evaporation of water is ubiquitous in nature and industrial technologies. The known mechanism for evaporation is "thermal evaporation", which highlights that the energy input for evaporation is via heat. Due to the weak absorption of water to visible light (*1,2*), the first step in using solar energy to evaporate water is usually by converting it into heat through photothermal processes via additional absorbing materials (*3-8*). Recently, we made the following unexpected experimental observations on evaporation from hydrogels under visible light illumination (*9*). (a) Partially wetted hydrogels become absorbing in the visible spectral range, where the absorption by both the water and the hydrogel materials is negligible. (b) Illumination of hydrogel under solar or visible-spectrum light-emitting-diode (LED) leads to evaporation rates exceeding the thermal evaporation limit, even in hydrogels without additional absorbers. (c) The evaporation rates are wavelength-dependent, peaking around 520 nm, despite that the absorptance does not show wavelength dependence in the same spectral range. (d) Temperature of the vapor phase becomes cooler under light illumination, and its distribution shows features different from thermal evaporation. And (e) vapor phase transmission spectra under the light show new features and peak shifts. We explained these observations by proposing that photons can directly cleave off water molecular clusters from the air-water interface without going through a photothermal process and call this effect the photomolecular effect, in analogy to the photoelectric effect discovered by Hertz (*10*) and explained by Einstein (*11*). This hypothesis also explains many past experimental observations of solar interfacial evaporation from porous materials exceeding the thermal limit of 1.45 kg/m$^2$-h obtained by converting 1000 W/m$^2$ (1 sun) of solar energy in one hour into the equivalent sensible and latent heat of water, assuming evaporation around 40 °C and an ambient temperature around 25 °C (*8,12-15*). However, evaporation from porous materials is difficult to quantify due to unknown internal structures. Here, we use a plethora of experimental methods to study visible-light (using mostly lasers) interactions with a single air-water interface. These experiments further support the proposed photomolecular effect.

In our previous work (*9*), we proposed the following possible mechanism of the photomolecular effect (Fig. 1a) to explain experimental observations on hydrogel. First, as shown in Fig. 1a, the density of water changes over a distance of a few angstroms (~3-7 Å) from the liquid to the vapor phase (*16,17*). Second, the incident light creates an electrical field gradient over the interfacial region. Based on the macroscopic Maxwell equations, the displacement field of the electromagnetic wave perpendicular to the interface is continuous at the interface, i.e., $\varepsilon_1 E_{1\perp} = \varepsilon_2 E_{2\perp}$, where $\varepsilon_1$=1 for air and $\varepsilon_2$=1.8 for water in the visible spectrum, and the subscript $\perp$ emphasizes the direction perpendicular to the interface (*18*). This condition means that the electrical field in the perpendicular direction is reduced by nearly a factor of two from air to water over a distance of a few angstroms, creating a large field gradient. Third, water molecules are polar and form fluctuating hydrogen bond networks, also called water clusters, although details of the networks are still under debate (*19-23*). The dipole moment of a single water molecular is 1.8D and increases with the cluster size (*24*), which means an effective charge separation of 0.5 Å or larger. Hence, the electrical field changes over the distance of a single water cluster at the interface is appreciable. Due to this variation, the force acting on the positive and negative charges of water molecules does not cancel out, leading to a net force exerting on the molecular clusters and pulling them from the interfacial region to the adjacent free space. This is similar in some ways to the surface photoelectric effect, also due to the large electrical field across 1-2 Å where the electron density changes across the interface (*25-27*). From the quantum picture of light, the interaction of a photon and a molecular cluster needs to conserve energy and momentum. The bond energy between one water cluster and its surrounding water molecules should be weakened, roughly between the normal hydrogen bonds (~0.26 eV) and the van der Waals bond (~0.026 eV). A green photon with an energy of ~2.5 eV can meet the energy conservation requirements via cleaving off many such bonds. Momentum conservation is satisfied because of the rapid change of the electrical field in the perpendicular direction, as is the case for the surface photoelectric effect (*26*).

Sum-frequency spectroscopy is an established method to study light interaction with a liquid interface (*28*). However, such nonlinear process requires high laser power. IR spectroscopy and Raman spectroscopy had been used to study water clusters produced in well controlled conditions (*29-32*). Our aim is to show that photomolecular effect happens at the air-water interface even under low light intensities comparable with solar radiation, for which there are no established methods. We probe photomolecular effect by studying the thermal and optical responses in the bulk water and the vapor phases, through measuring their polarization, angle, and wavelength dependence and directly observing the spectra signatures of clusters in the infrared region.

**Liquid-water surface temperature response.** We used an IR camera (Figs. 1b&1c) and a thermocouple (Fig. 1d) to measure the temperature response of the water surface induced by a green laser under different angles of incidence and polarizations (SM1.4&SM2, Table S1). Fig. 1b illustrates the water surface temperature response with time measured by the IR camera with the incident angle of 45° for both the transverse magnetic (TM) and transverse electrical (TE) polarized lasers. For the TM-polarized laser the water surface temperature increases by 0.7 °C in the first 30s after the laser turns on. It reaches a steady state with a rise of 0.95 °C after ~200 s. In contrast, the temperature rise under the TE-polarized laser is less than 0.3 °C.

The water surface temperature rises measured by both IR camera (Fig. 1c) and thermocouple (Fig. 1d) are consistently higher under the TM-polarized laser than under the TE-polarized beam. Furthermore, the temperature rise shows strong incident angle dependence, peaking at ~45° under the TM-polarized laser beam; in contrast, the temperature rise is nearly zero at normal incidence. Fig. S2 shows power dependence of the temperature rise at 45° incident angle, which is approximately linear.

**Vapor-phase temperature distribution above an evaporatively cooled surface**. We compare in Fig. 2a the vapor-phase temperature distributions with and without laser illumination, measured with a thermocouple (black circles and blue triangles) and an IR camera (black and blue line) (Fig. S3, SM1.5&SM4). Without laser illumination, water is cooled down from room temperature (22.8 °C) due to natural evaporation, leading to a temperature gradient from air to water. However, we observe that when the TM-polarized green-laser with an incident angle of 45º is turned on, the temperature of the vapor phase (Fig. 2a, Fig. S4) at 2 mm above the air-water interface decreases from 21.9 °C to 21.4 °C, while the water surface temperature increases from 20.7 °C to 21.3 °C. This phenomenon can be interpreted according to our hypothesized photomolecular effect (Fig. S5). Visible light can cleave off water clusters from the air-water interface and inject them into the air. Some of these clusters break up, absorbing heat and cooling down the air. Since the air convection is small, most of the clusters will condense back to the water surface, releasing heat to increase the water surface temperature.

**Vapor-phase refractive index change under light.** We probe the refractive index change of the vapor phase above the air-water interface through the mirage effect (*33,34*) (Figs. 2b&2c, Figs. S6&S7, SM1.6&SM4). A green laser with a power of 1.4 W and an incident angle of 45° works as pump and a red pen laser (635 nm) works as probe (red dash line in Fig. 2b) with a small incident angle of ~0.033 rad relative to the water surface. The probe beam displacement is measured by a beam position sensing detector between pump off and 30s after pump on, which can be mapped into the refractive index change along the pathlength. The effective average change of the refractive index ($\overline{\Delta n}$) of the vapor phase across the water surface (~30 mm in diameter) under the TM-polarized pump laser (wavelength 532 nm, power 1.4 W, $1/e^2$ radius 7.5mm) is $-6.7\times10^{-4}$, $-5\times10^{-4}$, and $-2.5\times10^{-4}$ at 1 mm, 3 mm, and 5 mm above the air-water interface, respectively. In contrast, the $\overline{\Delta n}$ of the vapor phase is lower than the measurement resolution ($10^{-4}$) when the same pump laser is TE-polarized. Since the thermo-optic coefficient ($dn/dT$) of air is $-9\times10^{-7}$ (*35*), the large refractive index change of the air is not due to the air temperature change. We instead attribute $\overline{\Delta n}$ to the existence of water molecular clusters in air.

We also measured the beam deflection in the transmission configuration (Fig. 2d, Fig. S8, SM1.7&SM5). In this measurement, the pump laser beam exciting molecular clusters also serves as a probe to measure the beam deflection. The direction of the transmitted beam changes (Fig. S9) when the laser power is changed. Although this method is inspired by the past pump-probe studies of thermal conductivity of solid, in which a probe beam is used to detect the refractive index change and surface deformation caused by a pump laser-induced heating (*36-39*), we found that the calculated beam deflection assuming the same effects (SM5.2) is over two-orders of magnitude too small. For example, for a TM-polarized red laser at 45° incident angle to a glass containing water (air-water/glass/air) with an assumed absorptance of 1% (Fig. S9) at the interface, the calculated beam deflection assuming only thermal effects in the liquid is <0.12 μrad/mW (Figs. S10-12, SM5.2), while the measured value is 49.1 μrad/mW (Fig. S9). We attribute the measured beam deflection to mainly the air-side refractive index change, consistent with the above-discussed measurements using the surface mirage effect.

We use the normalized beam deflection (SM5.3) to evaluate the polarization, angle and wavelength dependence of the incident laser (Table S2) on the photomolecular effect at the air-water interface. As shown in Figs. 2e & 2f, Figs. S13 & S14, and Table S3 & S4, the TE-polarized beam (red circles) has a much smaller normalized transmitted beam deflection than the TM-polarized beam (black squares). Fig. 2e illustrates the normalized transmitted beam deflection of a green laser (520 nm), which maximizes at the incident angle of 45°, consistent with the water surface temperature rise shown in Figs. 1c & 1d. Fig. 2f and Fig. S14 illustrate the photomolecular effect is strongest for green light (520 nm) at different incident angles, consistent with previous evaporation experiments on hydrogels (*9*), which show the highest evaporation rate under green light.

**Vapor-phase temperature distribution above a heated surface.** Under the ambient condition, there is little air motion since the water surface temperatures are only slightly lower than that of the ambient with or without light. The molecular clusters recondense back, causing heating of water as shown in Figs.1b-d. We heat up the water surface electrically to 53 °C to create a vertical draft of air flow and compare the vapor-phase temperature distributions with and without green-laser illumination onto the water surface (Fig. 2g-h). Without the laser illumination, the vapor-phase temperature decreases continuously (Fig. 2g). The TE-polarized laser beam does not cause much change in the vapor-phase temperature distribution (Fig. 2g). However, under a TM-polarized laser (Fig. 2h, Fig. S15a), a flat temperature region is observed at the distance between 6 mm and 20 mm. After the light is turned off, the air near the water surface heats up. We interpret such observations as follows (Fig. 2i). The clusters cleaved-off by the TM-polarized light are carried away by the ascending air plume. The clusters dissociate and absorb heat, making the air cooler near the water surface (Fig. S16). The flat region is formed when the air becomes saturated with water. Further away from the water surface, the air temperature decreases again as more dry air drafted from the surrounding ambient reduces local relative humidity below 100%. Such a flat temperature regions were also observed above a hydrogel surface (*9*) and above an air-water interface (Fig. S17) when they are subjected to green LED illumination. We interpreted the long flat temperature region as resulting from the combination of large air molecular speeds needed to break the hydrogen bonds and the large number of hydrogen bonds (*9*).

**Vapor-phase spectral signatures.** The transmission spectroscopy of the vapor-phase is performed over a hot water surface (53 ºC) (Fig. 3a, Fig. S18, SM1.8&SM6). The green laser system is coupled to the sample chamber of a commercial UV-Vis-NIR spectrometer to excite the water molecular clusters. A $CaF_2$ lens is used to focus the light source of the spectrometer to obtain enough intensity and spatial resolution (diameter of the beam at the focal plane is ~1 mm). The beam waist is ~4 mm above the hot water surface. Fig. 3b shows the vapor phase transmission spectra in the region of 3600 $cm^{-1}$ to 4000 $cm^{-1}$ with and without the laser beam illumination (TM-polarized), which differ significantly from each other in both magnitude and peaks. Also plotted in the figure is the absorption spectrum of pure water vapor (blue line), i.e., isolated water molecules (*40*). Clearly, under the TM-polarized laser illumination, the absorptance increased significantly, especially in regions where previous works had reported cluster signatures (*29-32*). We attribute the additional absorption to the water molecular clusters cleaved-off from the water surface.

Next, we use a Raman spectrometer to probe the spectrum of the vapor phase above the hot water surface (53 °C). As shown in Fig. 3c, the 532 nm laser of the Raman system works as both the pump and the probe (mixture of TE and TM). We used a 50X long working distance lens with the laser focal plane 3 mm above the air-water interface (denoted as AS, Fig. S19, SM7). The same laser beam intercepts the air-water interface at a range of angles, cleaving off water molecular clusters (the pump beam), which are probed around the focal point in the vapor phase since the Raman spectrometer is confocal. Compared with the case when the laser beam is focused on the water surface (denoted as OS, Fig. S19), the Raman spectrum of the AS configuration (Fig. 3d) has a clear peak around ~3650 cm$^{-1}$, which represents single water molecule (*41*), confirming that part of the signals indeed come from the vapor phase, superimposed on unavoidable Raman signal from liquid water that is significantly stronger than the vapor phase. In addition to this peak, multiple broad peaks in the range of 3000 cm$^{-1}$ to 3900 cm$^{-1}$ appear, which are qualitatively consistent with the transmission spectrum in Fig. 3b.

Cluster peaks and the single water molecular vapor peak (~3650 cm$^{-1}$) are also observed in the Raman spectrum when we focus the laser beam ~0.15 mm above cold-water surface (20.5 °C) under natural evaporation using a 100X lens (working distance = 0.21 mm) (Fig. 3e), since the heated water surface height changes too rapidly to allow Raman data collection using 100X lens. Compared to the OS Raman spectrum obtained using the same lens, which is dominated by water, in the AS configuration, multiple peaks of water cluster between 2900 and 3900 cm$^{-1}$ appear, qualitatively similar to that of absorptance spectra of clusters reported in literature (*29-32*). The Raman spectrum with different heights of beam waist above the air-water interface also shows cluster peaks (Fig. S20).

The commercial Raman system does not allow easy probing of the polarization effect. We study the polarization dependence on a custom-built Raman system (Fig. 3f, Fig. S21, Table S5, SM1.10&SM8). The green laser is modulated at f (=200 Hz) by a mechanical chopper to mimic a square wave modulation for which there exists no 2f harmonics (Fig. 3g, Fig. S22). In actual measurements, the intensity of 2f harmonics is 3% of that of 1f. The laser cleaves off water molecular clusters into the vapor phase. The water molecular clusters then interact with the same incoming laser to generate Raman signals at 2f (2ω Raman signal), which are measured by a photodiode detector connected to a lock-in amplifier. Filters are placed in front of the photodetector to ensure that only signals with wavenumber >2130 cm$^{-1}$ are measured. At an incident angle of 45°, the intensity of 2f signal is proportional to the square of power, $P^2$, for the TM-polarized laser because both the Raman signal of an individual molecule and the number of clusters are proportional to the power of TM-polarized laser (black squares in Fig. 3h), while the 2f scattering signal is approximately linearly proportional to the power of the TE-polarized laser (red circles in Fig. 3h).

**Mass Change under Modulated Light.** To quantify the mass change caused by the photomolecular effect, we measured the deflection of a clamped beam hosting a water container in the middle with a pump-probe configuration as shown in Fig. 4a (SM1.11&SM9). A modulated green-laser directed at the air-water interface leads to a periodic change of the vibrational amplitude of the clamped beam, which we interpret as being caused by the weight change in the container. The vibration is measured by the deflection of a continuous wave probe laser (635 nm) directed at a location without the pump laser nor water (Fig. S23). Fig. 4b illustrates the theoretical frequency response of the vibration of the used PMMA clamped beam subjected to a periodic force modulation at the center (SM9.2). The modulated signal in the low frequency range (region yellow) is constant, which means we can use the beam bending caused by the mass change during natural evaporation (without light) as a calibration (1 g/V) to infer the mass change created by the modulated laser beam. Fig. 4c shows the beam deflection under TM-polarized laser beam (black squares) is over 5 times larger than that under the TE-polarized laser beam (red circles), while the signal of the dry container (blue triangle) without water is smaller than the measurement uncertainty at the resonance frequency. Using the flat region beam deflection of ~10 µV in the low frequency range under the TM-polarized green light, we estimate that the mass change is ~10 µg under the laser power of ~0.89 W and incident angle of 45°. This mass change corresponds to a water layer ~14 nm. Since the modulation frequency is much higher than the characteristic time needed to reach steady state (Fig. 1b), we cannot relate this mass change to a steady-state evaporation rate.

**Absorptance Estimation**. Quantitative determination of the absorptance has been difficult. We estimate the absorptance based on the measured water surface temperature as follows (Fig. S24, SM10). First, using the evaporation rate without light and the measured water temperature, we estimate the effective convective heat transfer coefficient at 7.96 W/m$^2$-K. Under 1.4 W TM-polarized green light illumination at 45º, the water surface temperature increased by 0.54 °C (Fig. S25). Assuming the same heat transfer coefficient, the absorptance is estimated to be 0.84%.

**Superthermal Evaporation and Cluster Size Estimation.** Fig. 1 shows that most clusters recondense because the water surface is cold. To suppress the recondensation of the water clusters, we heat up the water to produce a rising plume (Figs. 4d, Fig. S26, SM1.12) or blow air across the cold-water surface (Figs. S27&28, SM1.12) to carry away the cleaved clusters so that the light-induced evaporation can be directly measured. Here, we use an IR lamp to heat up water to 33 ºC to avoid the possible absorption of the laser excitation when using an electrical heater. With the IR lamp power-on, a TM-polarized green laser (power/area of water surface≈1400 W/m$^2$) incident at 45º increases the evaporation rate by 0.07 kg/m$^2$-h while the TE-polarized light does not lead to any measurable change (Fig. 4d). We also measured a similar evaporation rate increase under a TM-polarized green laser by flowing a gentle wind across the cold-water surface (Fig.

S28). Assuming only 0.84% light is absorbed, the value of 0.07 kg/m$^2$-h would translate into 5.95 kg/m$^2$-h under 1000 W/m$^2$ incident light (power/area of water surface) if all light is absorbed using multiple interfaces. This is ~4 times of the thermal evaporation limit.

The energy needed to evaporate a water molecular is 0.46 eV. The photon energy at 532 nm is 2.33 eV. Assuming all the photon energy is used to break up water molecules, one green photon at this wavelength can evaporate 5.06 water molecules. Using the estimated evaporation rate of 5.95 kg/m$^2$-h for the photomolecular process under 1000 W/m$^2$, the average number of water molecules cleaved by a green photon is 20. These estimations on the evaporation rate and cluster size assume that all cleaved water clusters are brought away from the surface. In reality, some of the clusters may still condense back, and hence the actual values can be higher.

**Cloud Heating.** For a flat surface, the photomolecular absorption can be maximized by orienting the light to an optimal angle of incidence. For droplets, there are always electrical field directions of the light that are perpendicular to the interface. The increased surface area, the curvature of the droplets, and the multiple scattering effects will further enhance the photomolecular effect. Thus, we expect that photomolecular effect will be important for clouds and fogs. We designed and fabricated a cloud chamber and generated water droplets into the chamber, as shown in Fig. 4e (Fig. S29, SM1.13). LEDs of different wavelengths with intensities of 1000 W/m$^2$ were shone on the cloud chamber, and the temperature rise of an empty chamber with a cloud inside were measured using an IR camera. Fig. 4f show that the chamber itself is not heated up, but visible LEDs heat the cloud significantly, peaking at 520 nm, consistent with the steady state. Past cloud measurements have suggested significantly higher cloud absorptance than what existing models can predict (*42-44*), which had been subjected to questions due to difficulties in the experiment (*45,46*). The photomolecular effect and heating data we demonstrate here can provide both the theoretical basis and experimental support for increased solar radiation by clouds.

**Discussion.** In Table S6, we summarize 13 different experiments that support the existence of photomolecular effect: the cleavage of water molecular clusters by photons. These experiments cross-check different manifestations of the effect: polarization, angle of incidence, and wavelength dependence. Surprisingly, this effect happens at a spectral region where water is least absorbing. The qualitative pictures presented in Fig. 1a can reasonably explain the dependence on polarization and angle of incidence, but not the wavelength dependence. The wavelength dependence and the low intensity of light used suggest that the photomolecular effect is a single photon process, related to the matching of the photon energy to the water cluster binding energy with surroundings. However, we recognize that a quantitative description of the process will need more effort, especially considering that the states of the hydrogen bond network in bulk water and on the surface are unsettled (*22,28,47*). Past work on laser-induced desorption and surface photoelectric effects can be sources of inspiration for more theoretical studies (*27,48*).

We had proposed (*9*) that the photomolecular effect is the reason behind recent reports of solar-driven interfacial evaporation from different porous materials exceeding the thermal evaporation limit (*8,12-15*). We believe that the same mechanism could also explain some past experimental observations of light-induced breathing mode of water molecules on solid surfaces (*49*) and unexplained cloud absorption of solar radiation (*42-44,46*). Our demonstration of the photomolecular effect in fog suggests that this process is ubiquitous: from clouds to fogs, ocean to soil surfaces, and plant transpiration. We are wondering if plants are green, i.e., avoiding the most intensive spectrum of solar energy, have anything to do with the photomolecular effect, from evolutionary consideration. After all, biological organisms emerged way after the earth has water and sunlight. We often hear from weather forecasts "after the sun thinned the fog". Before, we thought this was caused by thermal evaporation. Our work shows that photomolecular effect could also be at play in this natural phenomenon. The stronger surface absorption of green light adds another mechanism to the beautiful color of sunset in addition to Rayleigh's explanation based on scattering. While it is well-known that bulk materials' absorption mainly happens via electronic and vibrational transitions, our work suggests that interfacial photomolecular effect can lead to photon absorption in water and possibly in other liquids or even solids. Further exploration of the effect can lead to new scientific insights and technologies in desalination, waste-water treatment, drying, and air conditioning.

**Acknowledgments:** We would like to thank Dr. Qichen Song for his help in optics and discussion, Dr. Shaoting Lin for coating hydrophilic hydrogel inside fog chamber, Carlos Daniel Diaz Marin, Professors Bolin Liao and Yangying Zhu for discussion and their attempts to repeat some of our experiments and Bolin Liao for comments on the manuscript, Dr. Alex Maznev for discussion on mirage effect, Dr. Rohith Mittapally, Dr. Qian Xu, Ms. Caterina Grossi and Ms. Briana Cuero for discussion and proofreading. We are also grateful to discussions with Professors Tonio Buonassisi (MIT), Ian Hunter (MIT), Kris Kempa (Boston College), Keith Nelson (MIT), Ron Shen (UC Berkeley), John Pendry (Imperial College), and Ely Yabnolovitch (UC Berkeley), although these discussions do not indicate of their endorsement. G.C. grateful thank MIT for its support and Tracy Chen for listening to his science and mentioning "when the sun thins the fog" that inspired the fog experiment.  This work started in 2021, and was partially supported by an MIT Bose Award since Fall 2022.

**Author contributions:**

Conceptualization: G.L., Y.T., and G.C.

Methodology: G.L., Y.T., and G.C.

Investigation: G.L., Y.T., J.Z., and G.C.

Visualization: G.L., Y.T., and G.C.

Funding acquisition: G.C.

Project administration: G.C.

Supervision: G.C.

Writing – original draft: G.L., Y.T., and G.C.

Writing – review & editing: G.L., Y.T., J.Z., and G.C.

**Competing interests:** Authors declare that they have no competing interests.

**Data and materials availability:** All data are available in the main text or the supplementary materials.

**Supplementary Materials**

Materials and methods

Supplementary Text

Figs. S1 to S33

Tables S1 to S6

References (50-52)

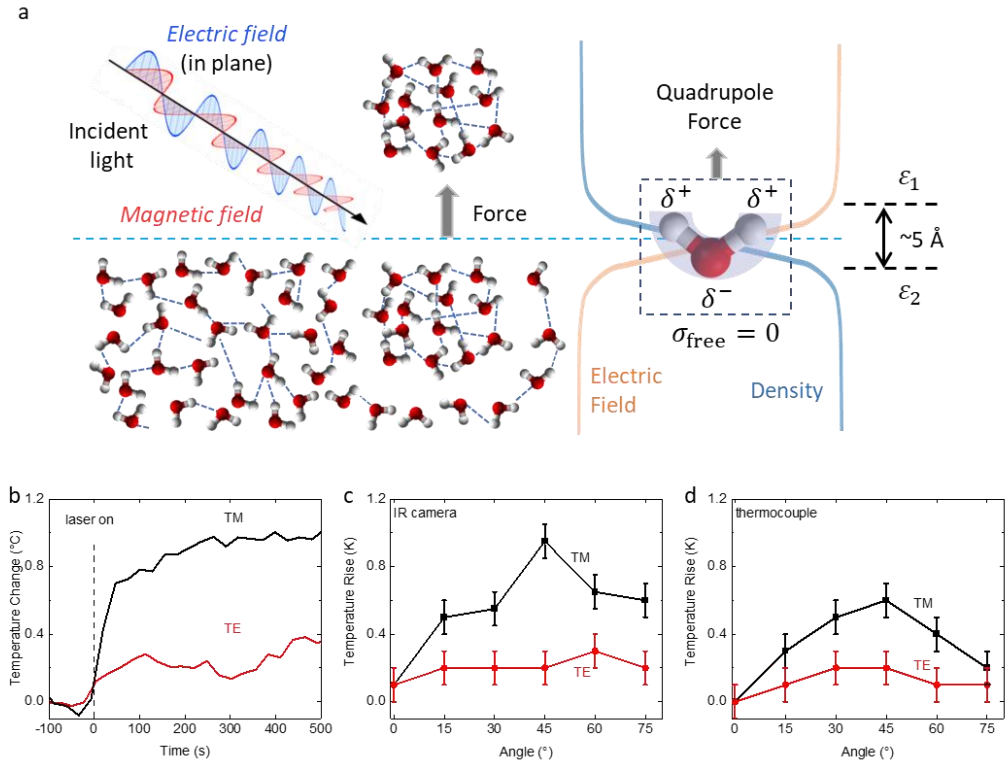

**Fig. 1. Proposed photomolecular effect and measured water temperature rise under laser illumination.** (a) Photomolecular effect happens when a TM-polarized laser shines on the air-water interface at an angle such that an electrical field component perpendicular to the surface changes rapidly across the interface, leading to a net force acting on polar water molecular clusters. The clusters are driven out of the air-water interface when the photon energy matches the required energy to cleave off the cluster from its surroundings. (b) Temperature rise of the water surface as a function of time after a 532 nm green laser at 45º incident angle is turned on, measured by an IR camera. Time zero is defined when the laser turns on. The surface water temperature rise under the TM-polarized laser is much larger than under the TE-polarized laser. (c)&(d) Temperature rise as a function of incident angle measured by (c) an IR camera and (d) a thermocouple, shows consistency between the two, with a peak at 45º under the TM-polarized laser. The power of laser is 1.4 W. The $1/e^2$ radius of the laser is 7.5 mm. The radius of the air-water interface is 15 mm.

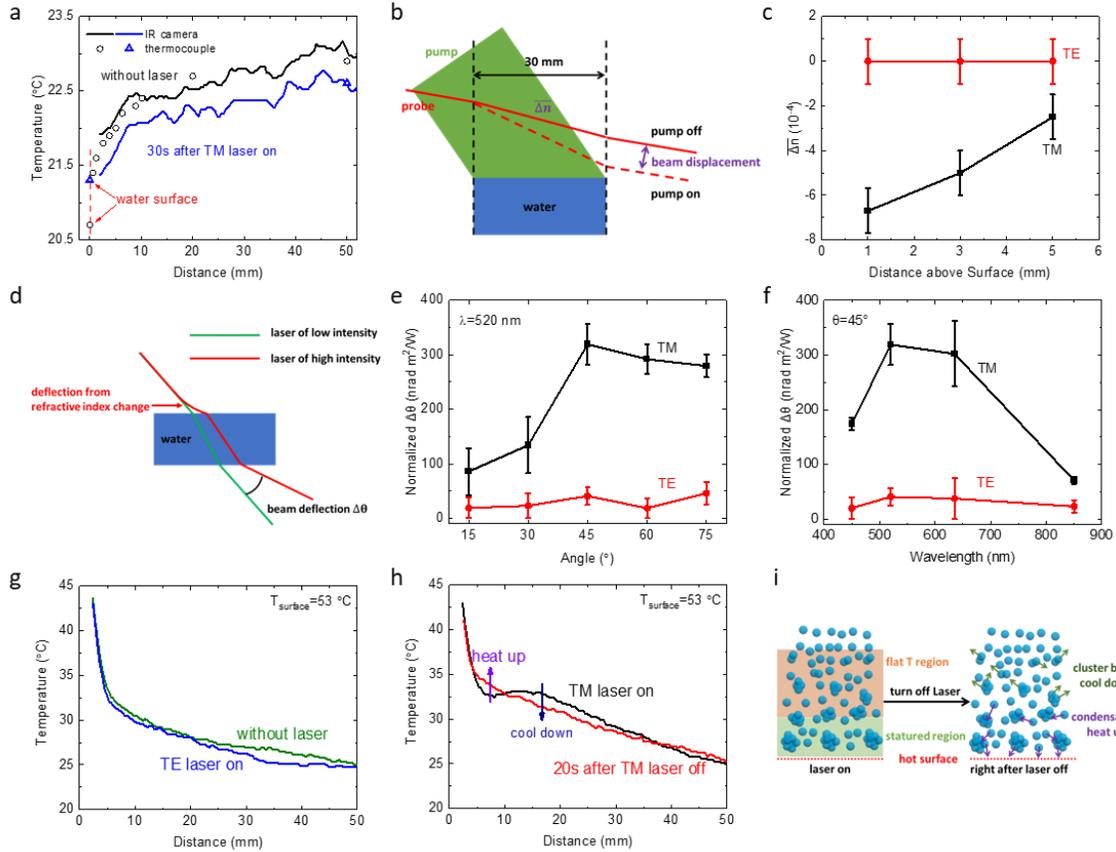

**Fig. 2. Vapor phase temperature and refractive index responses.** (a) Vapor-phase temperature distribution above the air-water interface before and after TM-polarized laser illumination (wavelength 532 nm, power 1.4 W, incident angle 45°, $1/e^2$ radius 7.5 mm). The radius of the air-water interface is 15 mm. Solid lines (black when laser off, blue when laser on) measured by an IR camera, circles (laser off) and triangles (laser on) measured by a thermocouple. With the laser on, the air temperature drops while the water temperature increases. (b) Schematics of the mirage effect experiment: a probe laser beam (635 nm) aligned ~0.033 rad off-parallel from above the air-water interface deflects due to the refractive index gradient in air created by green laser illumination as in (a). (c) Effective average refractive index change as a function of the probe beam distance above the air-water interface for the TM-polarized and TE-polarized green pump laser. (d) Schematics of steady-state transmitted beam deflection measurement: pen lasers of different wavelengths and angles of incidence are directed towards air-water interface. Deflection of the transmitted beam is measured at different laser power. The deflection is caused mainly by the refractive index change in the air due to cluster excitation. (e) Normalized beam deflection as a function of the incident angle of a green pen laser (520 nm) for TM (black squares) and TE (red circles) polarizations. (f) Normalized beam deflection as a function of wavelength for TM and TE polarizations at 45˚ incident angle. (g)&(h) Comparison of vapor phase temperature distributions measured away from the air-water interface when water is initially heated to 53 ºC with and without laser (wavelength 532 nm, power 1.4 W, incident angle 45°, $1/e^2$ radius 7.5 mm). (g) No difference is observable with and without TE-polarized green laser (h) For TM-polarization, a flat temperature region is observed, significantly different from the vapor temperature profile 20 s after the laser turns off. (i) Explanation for the vapor phase temperature distribution of hot water: under TM-polarized laser illumination, clusters cleaved off move away from surface with rising plume, break up in air and saturate the air, creating flat temperature region when air is saturated.

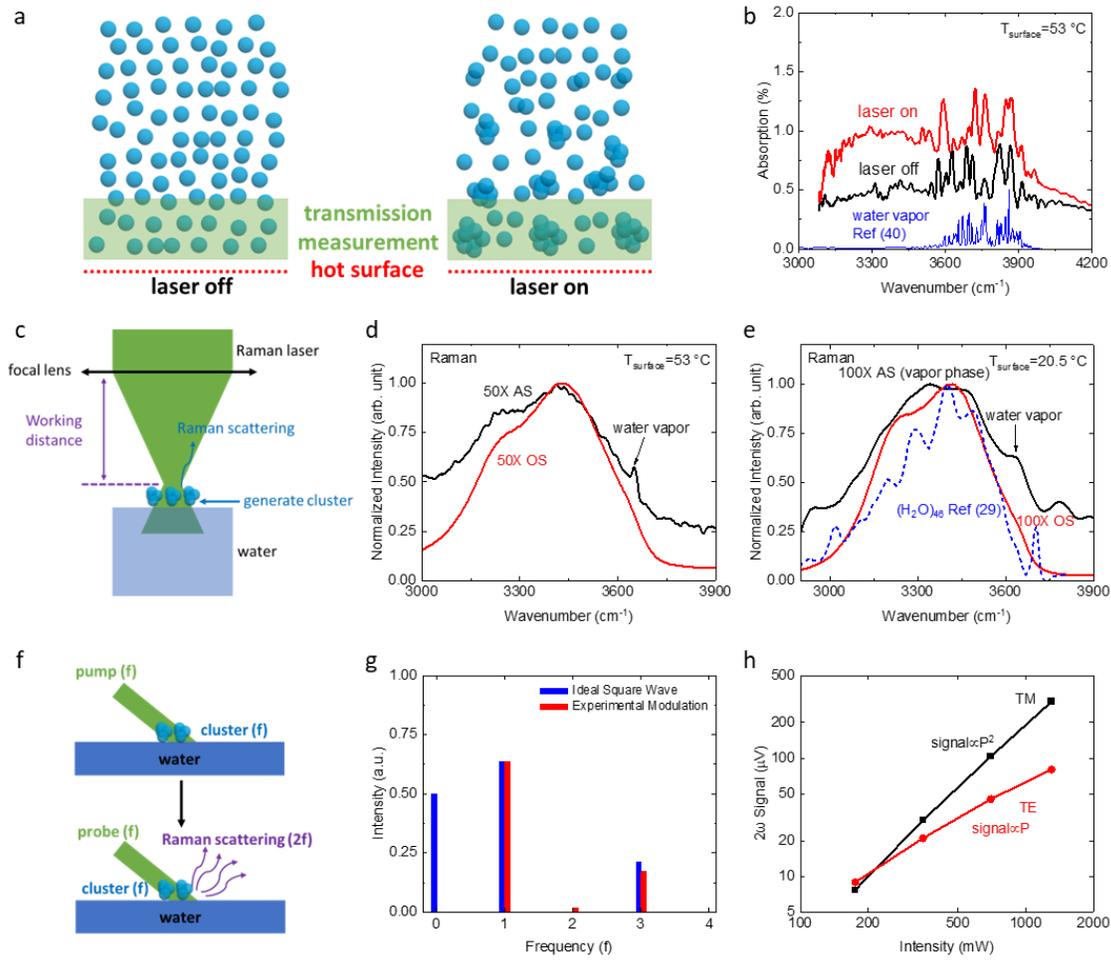

**Fig. 3. Spectra of vapor phase.** (a) Illustration of IR transmission measurement above electrically heated water surface (53 ºC, 30 mm in diameter) with and without green laser illumination (532 nm, power 1.4 W, $1/e^2$ radius close to air-water interface ~9 mm, incident angle 40-50º, TM-polarized).The radius of air-water interface is 15 mm. IR probe beam is condensed to ~1 mm in diameter and is ~4 mm above air-water at the beam waist. (b) Absorptance spectrum of the vapor phase with green laser on and off. Blue line represents the absorption of pure water vapor (*40*). (c) Raman spectrum measurement of vapor phase. When the focal point of the excitation laser beam (532 nm) is above the air-water interface (AS), the divergent excitation laser after the focal point hits water surface, generating clusters. (d) Raman spectra with focal point ~ 3mm above (AS) and on (OS) the hot water surface (53 °C before laser turns on), probed with a 50X lens (33 mW, 26.5 mm working distance, beam waist $1/e^2$ radius ~1 μm). The OS spectrum is consistent with water and AS spectrum contains clear features of water vapor, together with other peaks indicative of clusters existence. (e) Raman spectra of unheated air-water interface (~20.5 ºC before laser is turned on) probed with a 100X lens (0.21 mm working distance, 17 mW, beam waist $1/e^2$ radius ~0.5 μm). The OS spectrum is consistent with water and the ~0.15 mm AS spectrum contains peaks resemble that of the IR spectrum of 46-water cluster vapor reported in literature (*29*). (f) Schematic of custom-built Raman scattering measurement: a polarized laser beam modulated by a mechanical chopper into a square wave at fundamental frequency ω=1256 Hz (f=200 Hz) is directed towards water surface (532 nm, 0.75 W, $1/e^2$ radius ≈ 2.8 mm, incident 45º), cleaving off clusters into air.

These clusters interact with the same laser beam. The scattered Raman signals at $2\omega$ are collected by detector after an optical filter. (g) Fourier series of a square wave, showing that an ideal square wave has no (blue) and the actual modulated beam (red) with weak $2\omega$ component. (h) $2\omega$ Raman signal as a function of the laser beam power, displaying square and linear law dependence for TM and TE-polarized beam, respectively. The square law supports clusters in air, since both the Raman signal and the number of clusters are proportional to power.

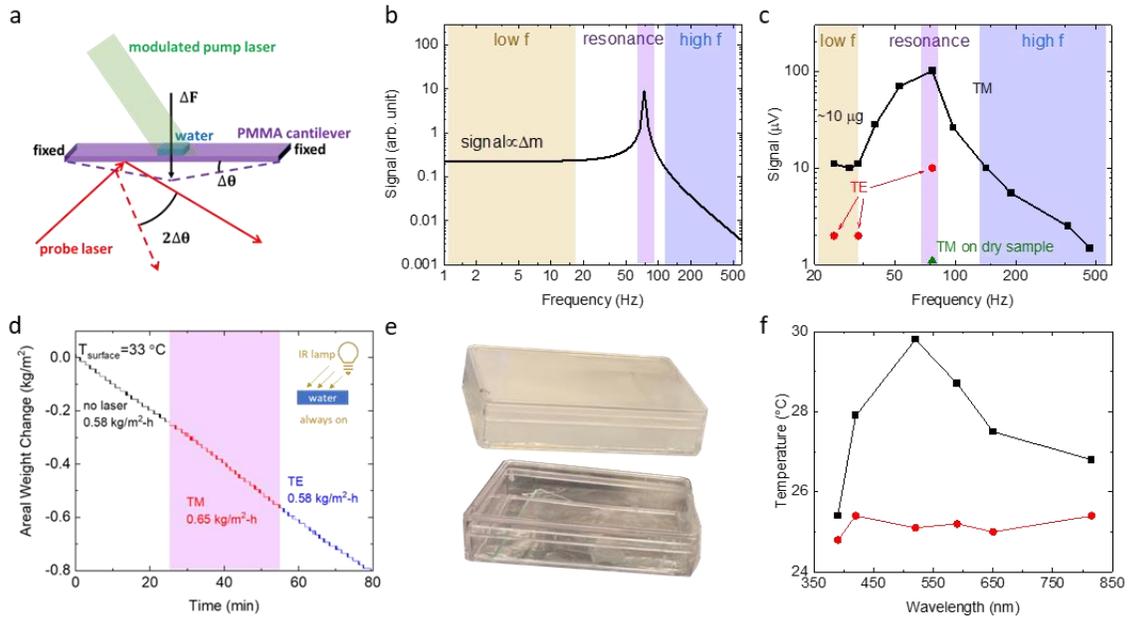

**Fig. 4. Modulated pump-probe and frog measurements.** (a) Schematic of modulated clamped beam (MCB) measurement. The container of water is not displayed in the figure. (b) Theoretical beam deflection signal of PMMA bending cantilever as a function of frequency. (c) Measured beam deflection signal as a function of modulation frequency. Black squares and red circles represent TM-polarized and TE-polarized pump laser on water, respectively. Green triangle represents TM-polarized pump laser on dry container without water. (d) Weight changes (normalized to area) measured by balance as a function of time initially heated by an IR lamp to 33 °C, followed by TM-polarized and TE-polarized laser illumination (1.0 W). The radius of air-water interface is 15 mm. Power/area of water surface ≈ 1400 W/m$^2$ (1.4 suns). (e) Photos of the fog chamber with and without fog. (f) The steady state temperature of the clouds in a chamber under different LED illumination (1000 W/m$^2$). The black squares and red circles represent the temperature of the fog chamber with fog and without fog, respectively.

# Supplementary Materials for

# Photomolecular Effect: Visible Light Interaction with Air-Water Interface

Guangxin Lv[1], Yaodong Tu[1], James H. Zhang, Gang Chen*

**Affiliations:**

Department of Mechanical Engineering, Massachusetts Institute of Technology, Cambridge, MA 02139.

*Corresponding author. Email: gchen2@mit.edu.

[1]These authors contributed equally to the work.

**This PDF file includes:**

    Materials and Methods

    Supplementary Text

    Figs. S1 to S33

    Tables S1 to S6

    References (50-53)

**Supplementary Text**

## SM1. Materials and Method

### SM1.1 Materials and container

Deionized water with resistivity 0.182 MΩm is used. The outer diameter, the inner diameter and height of the glass container are 33 mm, 30 mm, and 10 mm, respectively. The thickness of the bottom glass of the glass container is ~1 mm. The outer diameter and height of polystyrene (PS) container is 36 mm, and 12 mm, respectively.

### SM1.2 Laser and LED

The parameters of the pen lasers used (from Thorlab) are summarized in Table S2.

1.5 W high power green laser (from OPTO Engine LLC) has $1/e^2$ diameter ~2.0 mm. The expansion of the laser is done by two focal lenses with different focal lengths (Fig. S1). Incident angle is changed by two mirrors (Fig. S1).

LED lamps (from Charon) used have a rated power of 100 W and different nominal wavelengths: purple 390 nm, blue 440 nm, green 520 nm, yellow 590 nm, red 650 nm, and IR 815nm.

OT301 precision position sensing amplifier was purchased from ON-TRAK.

### SM1.3 Thermocouple and IR cameras

K-type thermocouple (diameter = ~0.15 mm) from Omega was shaped into a circle with diameter of ~8 mm. IR cameras used in this work are FLIR C5 and FLIR ETS320. In IR camera measurement, the emissivity of water is calibrated by adjusting the reading of water at the temperature of 20.7 °C (measured by a thermocouple). The emissivity of coverslip and PS are calibrated by adjusting the reading of coverslip and PS at the ambient temperature 22.8 °C (measured by a thermocouple). The humidity and ambient temperature are read by a humidity meter (Mengshen, M86).

**SM1.4 IR image & thermocouple measurements of water surface temperature**

The optical path is shown in Fig. S1. The power and $1/e^2$ radius of the expanded high power green laser (532 nm) are 1.4 W and 7.5 mm, respectively. The glass container was placed ~30 cm above the optical table (Fig. S30) with transparent support (glass slide and PMMA) to avoid heat transfer with the optical table heated by the transmitted laser. The steady-state temperature rise of dry glass container is <0.1 °C under the same laser power measured by an IR camera. The experiment was performed in a quiet laboratory environment and a shield was used to decrease the influence of wind on the steady-state evaporation rate. Under the relative humidity (HD) ~40% and environment temperature of 22.8 °C, the steady-state temperature of water surface is ~20.7 °C and natural evaporation rate of water is ~0.09 kg/m$^2$-h. It often takes 1 h for the water at room temperature to reach the steady state.

**SM1.5 Vertical vapor phase temperature distribution**

Optical path for measuring the temperature profile of the vapor phase above the air-water interface is shown in Fig. S3. The power and $1/e^2$ radius of the expanded high power green laser (532 nm) are 1.4 W and 7.5 mm, respectively. The polarization of pump laser beam was changed by a half-wave plate. A coverslip was used as a thermal emitter to indicate the vapor phase temperature since the vapor itself does not emit much thermal radiation (Fig. S31). The coverslip is thin (0.1 mm thick) which minimize heat conduction along the height direction and is suspended

~2 mm above water surface (not touch water surface) by a thin string made of cotton to minimize heat conduction distortion of the temperature distribution. An IR camera (FLIR) was used to measure the temperature of the coverslip.

For heated water, Joule heating (resistance wire glued to the bottom of the glass container) is used to heat up the water. For room temperature evaporation, no heater was used with the glass container.

**SM1.6 Vapor-phase mirage effect measurement**

Optical path of mirage effect of vapor phase above air-water interface is shown in Fig. S6. The power and $1/e^2$ radius of the expanded high power green laser are 1.4 W and 7.5 mm, respectively. A pen laser (635 nm) was used as the probe beam, which is arranged perpendicular to the plane of incidence of the pump beam. The probe laser was adjusted to have an angle with water surface (~ 0.033 rad) and the beam path is adjusted to the height 1, 3, and 5 mm above air-water interface. A position sensing detector connected to OT301 precision position sensing amplifier (ON-TRAK) was used to measure the shift of probe beam position before and 1-minute after the pump laser turned. A 600 nm long pass filter was placed before the position sensing detector to avoid the influence of scattered pump laser on the measurement of beam position shift. In reality, we found that the set-up with and without the filter gives the same signal because the directions of pump and probe beams are perpendicular.

**SM1.7 Steady-state transmitted beam deflection (SS-TBD)**

The set-up of the SS-TBD measurement is shown in Fig. S8. Pen lasers (parameters shown in Table S2) were used. We rotated the pen lasers to change the polarization of the laser instead of using a half-wave plate since the half-wave plate will lead to the beam deflection when the power of laser is changing. A position sensing detector connected to OT301 precision position sensing amplifier (ON-TRAK) was used to measure the beam deflection. Since the beam position sensing detector needs laser input to function, we use the slope of the beam deflection vs power (Fig. S9) to quantify the beam deflection caused by the photomolecular effect.

**SM1.8 Transmission of vapor phase**

The optical path of the vapor phase transmission measurement is shown in Fig. S18. The green laser was added into commercial Cary Series UV-VIS-NIR spectrophotometer. Expanded high power green laser was used as pump laser. The polarization of pump was changed by a half-wave plate. Due to the limited space of the sample compartment of the commercial spectrophotometer, the pump-beam has some divergence. As a result, the incident angle is in the range of 40° to 50° (average is 45°), the polarization is a TM-polarization, and the power and $1/e^2$ radius of the expanded high power green laser close to air-water interface are 1.4 W and ~9 mm, respectively. These parameters are slightly different from other experiments in this work. CaF lens (IR transmitted, focal length = 2 mm) was used to focus the light source of the spectrometer to get enough intensity and spatial resolution (diameter of beam at focal plane is ~1 mm). The air-water interface was adjusted and placed at ~4 mm below the beam waist.

**SM1.9 Raman spectroscopy**

A commercial confocal Raman spectrometer (HORIBA Scientific) was used to measure the spectrum of liquid water and vapor by varying the position of the focal plane (Fig. S19). The wavelength of the laser in Raman system is 532 nm.

In a confocal Raman system, the signal detected by CCD camera is weighed Raman scattering of all components close to the beam waist (Fig. S19). We catalog the configurations into 4 groups based on the position of beam waist/focal plane: on the water surface (OS), above the water surface (AS), below the water surface (BS), and emersion lens (EL). OS and BS are common configurations used in most of the past Raman studies of water. In our Raman system, the $1/e^2$ radius at the back focal plane of the objective lenses is ~1.1 mm.

**SM1.10 2ω Raman scattering measurement.**

The optical path of the 2ω Raman scattering measurement is shown in Fig. S21. The high power green laser (1.4 W) was modulated by a mechanical chopper (modulation frequency f=200 Hz) placed at the focal plane of the two objective lenses to be closer to an ideal square wave. Its polarization was changed by a half-wave plate. The power and $1/e^2$ radius of the expanded high power green laser are 0.7 W and 5.7 mm, respectively. The mechanical chopper provides a reference signal (f) to the lock-in amplifier. A 550 nm long pass filter and a 600 nm long pass filter are placed between the sample and photodiode detector to filter the elastic scattering of green laser. The photodiode was placed 1 cm away from the beam spot on surface and not directly exposed to the incident and reflected laser. The photodiode was connected to the lock-in amplifier.

**SM1.11 Measurement of modulated clamped beam (MCB)**

The optical path is shown in Fig. S23. The expanded high power green laser works as the pump. The power and $1/e^2$ radius of the expanded modulated high power green laser are 0.7 W and 7.5 mm, respectively. The polarization of the pump laser was changed by a half-wave plate. The pump beam was modulated by a mechanical chopper, which provides a reference signal to the lock-in amplifier. A pen laser (635 nm) was used as probe, which is perpendicular to the plane of incidence of the pump laser. A 600 nm long pass filter was placed before the position sensing detector to avoid the influence of scattered pump laser on the measurement of the beam position shift, although we obtained the same signal with and without filter because the pump and probe beams are perpendicular to each other. The beam position sensing detector was connected to a lock-in amplifier to measure modulated beam deflection signal. Time constant of the lock-in amplifier is set to 30 s for f>100 Hz and 100 s for f<100 Hz. The length (L), width (W), and thickness (D) of PMMA beam are 150 mm, 20 mm, and 0.25 mm. These values were designed considering the amplitude of the signal, and the noise, and the modulation frequency range. The water container was put in the middle of the PMMA beam. A mirror was stuck under the bottom of the cantilever and not exposed to pump beam, shown in Fig. S23.

**SM1.12 Measurement of evaporation under an IR lamp or air flow**

Optical paths of the evaporation measurement under an IR lamp and air flow are shown in Fig. S26 and Fig. S27, respectively. Air flow is produced by a fan. The power and $1/e^2$ radius of the expanded high power green laser are 0.7 W and 7.5 mm, respectively. The surface of balance surface is covered with highly reflective mirror film (3M ESR reflector film) to avoid the heating of balance under laser illumination.

**SM1.13 Fog generation and temperature measurement**

The fog generation system is illustrated in Fig. S29. Here the fog generator is made of an ultrasonic Atomization Maker (20 mm in diameter, 113 kHz, from WHDTS@Amazon). To supply water and prevent the generator to heat up, we built a water-cooling kit by using a silicone tube and plastic tube. The water-cooling kit was glued to the backside of the ultrasonic Atomization Maker. The fog size according to the product manual is about 3 micrometers in diameter. To avoid dropwise condensation on the chamber surface which blocks the light transmission, we coated a thin layer of hydrogel on the inside surface of the chamber (*50*).

To show the fog can absorb visible light due to the photomolecular effect, we used LEDs lamps with various wavelengths as the light sources and an IR camera to monitor the fog chamber's surface temperature.

The testing sequence consists of the following steps: (a) leave the dry and empty fog chamber on the test platform for 60 minutes, and then monitor its outside top surface temperature for 20 minutes to make sure it gets thermal equilibrium with the surrounding environment; (b) turn on the LED lamp and shine onto the fog chamber (empty without fog) for 60 minutes, and then monitor the outside top surface temperature of the dry and empty fog chamber for 20 minutes to make sure it gets thermal equilibrium with the surrounding environment. This step makes sure no absorption of the chamber. The chamber does not heat up under LED; (c) turn off the LED lamp and turn on the fog generator, after 30 minutes when the chamber is filling with fog (Fig. 4e), start to monitor the outside top surface temperature of the fog chamber for 20 minutes to make sure it gets thermal equilibrium with the surrounding environment; (d) turn off the fog generator, after 30 minutes when almost all the inside fog condense onto the wall, turn on the LED lamp for 60 minutes, and then monitor the outside top surface temperature of the fog chamber for 20 minutes to make sure it gets thermal equilibrium with the surrounding environment. In this case, we can make sure no obvious temperature rise due to the absorption of the condensate film; (e) drain out the condensation inside the fog chamber and turn on the fog generator, after 30 minutes when the chamber is filling of fog (Fig. 4e), turn on the LED lamp after 30 minutes start to monitor the outside top surface temperature of the fog chamber for 20 minutes to make sure it gets thermal equilibrium with the surrounding environment. For all the measurements under various LED lamps, we keep the light intensity of 1000 W/m$^2$ (1 sun) on the container surface and the same distance between the lamp and the fog chamber.

## SM2. Water Surface Temperature Measurement under Green Laser Illumination

IR camera and thermocouple are used to measure the temperature rise at the air-water interface. The optical path is shown in Fig. S1. A high power 532 nm CW laser was utilized to create enough temperature rise to meet the temperature resolution of the IR camera and the thermocouple (0.1 °C). Objective lenses with focal lengths of 20 mm and 150 mm were used to expand the size of the green laser. The power and $1/e^2$ radius of the expanded high power green laser (532 nm) are 1.4 W and 7.5 mm, respectively. The radius of water surface is 15 mm. The glass container was suspended (Fig. S30) and placed at a height of 30 cm away from the optical table to avoid the influence of heat exchange with the optical table. Two mirrors were utilized to change the incident angle of the laser. No temperature change is observed on the glass container (without water) with laser off and laser on using the same expanded high power green laser.

Since the laser influences the temperature reading of the thermocouple, the thermocouple is parked away from the optical path during laser illumination. The steady-state temperature of water surface under laser is measured ~5s seconds after the laser turns off---the time took to use the micrometer to move the thermocouple to the same position where the steady-state temperature before laser on was measured.

Since the laser beam is Gaussian and is not large enough to blanket the water container in all experiments, it often intercepts the water contained as shown in Fig. S32. In angle dependence experiments, we do not vary the laser power. The intensity distribution of the laser beam on the water surface in a Cartesian coordinate is

$$I(x,y) = \frac{2P_e * \cos(\theta)}{\pi w_e^2} \exp\left(-\frac{2\sqrt{x^2+y^2\cos(\theta)^2}}{w_e^2/\cos(\theta)}\right) / \sqrt{x^2+y^2} \tag{S1}$$

where $P_e$ is the power of expanded laser (1.4 W), $w_e$ is the $1/e^2$ radius of expanded laser (7.5 mm), $\theta$ is the incident angle.

The power of the laser intercepted at the air-water interface ($P_{interface}$) can be calculated by

$$P_{interface} = \int_{-r}^{r} \int_{-\sqrt{r^2-x^2}}^{\sqrt{r^2-x^2}} I(x,y) \, dy \, dx \tag{S2}$$

where r is the radius of air-water interface (15 mm). The power of laser on air-water interface is summarized in Table S1.

We can normalize $P_{interface}$ to the container area to obtain the power density at the air-water interface. Dividing this power density by $\cos\theta$ gives the equivalent incoming light intensity. These numbers are also given in Table S1.

## SM3. Measurement of Vapor Phase Temperature Distribution

The optical path for the IR measurement of the temperature profile of the vapor phase above the air-water interface is shown in Fig. S3.

A high power 532 nm CW laser was expanded by two objective lenses with focal length of 20 mm and 150 mm. The power and $1/e^2$ radius of the expanded high power green laser (532 nm) are 1.4 W and 7.5 mm, respectively. The radius of water surface is 15 mm. Two mirrors were utilized to change the incident angle of pump. The polarization of pump is changed by a half-wave plate.

The glass container was suspended and placed at a height of ~30 cm away from the optical table to avoid the influence of heat exchange from optical table (Fig. S31).

A thin glass coverslip (thickness 0.1 mm, length 50 mm, width 23 mm) was used to thermalize with the vapor so that its temperature can be measured with an IR camera (*9*). The glass coverslip was suspended ~2 mm above water surface (not touch water surface) by a thin string made of cotton to minimize the heat conduction loss from the glass slide. An IR camera (FLIR C5) was used to image the temperature profile of the coverslip.

**SM4. Measurement of the Vapor Phase Refractive Index Change Using Mirage Effect**

The optical path for probing the vapor phase refractive index change using the mirage effect of is shown in Fig. S6.

A high power 532 nm CW laser expanded by two objective lenses with focal lens of 20 mm and 150 mm were used as pump laser. The power and $1/e^2$ radius of the expanded high power green laser (532 nm) are 1.4 W and 7.5 mm, respectively. Two mirrors were utilized to change the incident angle of pump. The polarization of pump was changed by a half-wave plate.

A pen laser (635 nm) was used as probe, which is directed perpendicular to the plane of incidence of the pump laser. As shown in Fig. S7, the probe laser is adjusted to have a small angle $\theta_{probe}$ with horizontal plane (~0.033 rad or ~1.9º) and a position sensitive detector was used to measure the shift of beam position before and 1 minute after the pump was turned on. At this angle of incidence and over the 30 mm water container diameter (denote as $d$), the difference of the probe beam between the inlet and exit point above the water container is ~0.99 mm, while the laser $1/e^2$ beam radius in the height direction is ~0.64 mm. The beam displacement ($\Delta x$) is measured to be ~0.66 μm (TM-polarized pump and probe 1 mm above air-water interface). A 600 nm long pass filter was placed before the position sensitive detector to avoid the influence of the scattered pump laser on the measurement of beam position shift, although we found negligible difference between using and without such a filter because the directions of pump and probe beams are perpendicular to each other.

The effective average refractive index change ($\overline{\Delta n}$) in the region above water interface (diameter $d$ = 30 mm) is calculated from Snell's law

$$n \sin(\theta_{probe}) = (n + \overline{\Delta n}) \sin(\theta_{probe} + \Delta\theta_{probe}) \tag{S3}$$

Since $\overline{\Delta n} \ll n$, $\Delta\theta_{probe} \ll \theta_{probe}$, $\theta_{probe}$ is a small value, and $n = 1$, we get

$$\Delta\theta_{probe} = -\overline{\Delta n}\theta_{probe} \qquad (S4)$$

The beam displacement ($\Delta x$) follows

$$\Delta x = \Delta\theta_{probe} d \qquad (S5)$$

As a result, the refractive change is

$$\overline{\Delta n} = \frac{\Delta x}{\theta_{probe} d} \qquad (S6)$$

## SM5. Steady-state Transmitted Beam Deflection Measurement (SS-TBD)

**SM5.1 Measurement of SS-TBD**

The set-up of the SS-TBD measurement is shown in Fig. S8. Pen lasers of different wavelengths, including 450 nm (blue), 520 nm (green), 635 nm (red), and 850 nm (IR), were used without beam expansion. The parameters of these pen lasers were listed in Table S2. We rotated the pen lasers to change the polarization of the laser instead of using a half-wave plate since the half-wave plate will lead to beam deflection when the power of the laser is varied. Two mirrors were used to adjust the incident angle at the air-water interface. A position sensing detector connected to OT301 precision position sensing amplifier (ON-TRAK) was used to measure the beam deflection. Since the position sensitive detector needs input power to display the position of the laser, we measure the beam deflection under different laser power and extract the beam deflection per unit power in the analysis (Fig. S9, black curve). No beam deflection is observed on the glass container without water (Fig. S9, red curve).

In the SS-TBD experiment, the laser beam serves as both a pump and a probe. The laser beam can cleave off water clusters into the vapor phase, some of which will condense back creating heating in water, as we learnt from Fig. 1. Water heating can change the refractive index, leading to beam bending, or change the slope of local water surface curvature due to thermal expansion. The clusters in the vapor phase will also lead to refractive index change. In the Mirage experiment, we already learnt that the vapor side refractive index change is the dominant reason for the beam bending. Below, we will estimate the order of magnitude of the photomolecular-effect induced heating in air and water on the beam deflection.

**SM5.2 Model of thermal effect of SS-TBD**

To calculate the temperature profile with a Gaussian beam absorption on the air-water interface, a bidirectional transport model is used (*36*). The coordinate system is shown in Fig. S10. We assume photomolecular effect generates heat at the air-water interface, with $A_p$ denoting the absorbed laser power following

$$A_p = \alpha P \tag{S7}$$

where $\alpha$ is the absorptance and $P$ is the projected power of the laser ($=I\cos(\theta)S_{surface}$). $\theta$ is the incident angle, $I$ is the effective intensity of the laser, and $S_{surface}$ is the area of air-water interface.

The Hankel transform of the surface temperature rise of the layered structure by Gaussian shaped pump beam heating is

$$\Delta T(k) = G(k)P(k) \tag{S8}$$

where $P(k)$ is the Hankel transform of the pump beam intensity, which is defined by

$$P(k) = A_p \exp\left(-\frac{\pi^2 k^2 w_p^2}{2}\right) \tag{S9}$$

where $G(k)$ is a function of thermal parameter including thermal conductivity $\Lambda$, volumetric heat capacity $C$, the thickness of $d$ of water and air, as we show below.

The general model considers a heating source ($j = h$) at one interface of $N$ layers, which propagates up and down with G$_{up}$ and G$_{down}$ as the propagators. With thermally thick top and bottom layers, G$_{up}$ and G$_{down}$ can be expressed as

$$G_{up} = \frac{1}{\gamma_h}\left(\frac{A_h^+ + A_h^-}{A_h^- - A_h^+}\right) \tag{S10}$$

$$\begin{pmatrix}A^+\\A^-\end{pmatrix}_j = \frac{1}{2\gamma_j}\begin{pmatrix}e^{-\mu_j d_j} & 0 \\ 0 & e^{\mu_j d_j}\end{pmatrix}\begin{pmatrix}\gamma_j + \gamma_{j-1} & \gamma_j - \gamma_{j-1} \\ \gamma_j - \gamma_{j-1} & \gamma_j + \gamma_{j-1}\end{pmatrix}\begin{pmatrix}A^+\\A^-\end{pmatrix}_{j-1} \tag{S11}$$

$$\begin{pmatrix}A^+\\A^-\end{pmatrix}_1 = \begin{pmatrix}1\\0\end{pmatrix} \tag{S12}$$

$$G_{down} = \frac{1}{\gamma_{h+1}}\left(\frac{B_{h+1}^+ + B_{h+1}^-}{B_{h+1}^- - B_{h+1}^+}\right) \tag{S13}$$

$$\begin{pmatrix}B^+\\B^-\end{pmatrix}_j = \frac{1}{2\gamma_j}\begin{pmatrix}e^{-\mu_j d_j} & 0\\ 0 & e^{\mu_j d_j}\end{pmatrix}\begin{pmatrix}\gamma_j+\gamma_{j-1} & \gamma_j-\gamma_{j-1}\\ \gamma_j-\gamma_{j-1} & \gamma_j+\gamma_{j-1}\end{pmatrix}\begin{pmatrix}B^+\\B^-\end{pmatrix}_{j+1} \quad (S14)$$

$$\begin{pmatrix}B^+\\B^-\end{pmatrix}_1 = \begin{pmatrix}0\\1\end{pmatrix} \quad (S15)$$

where $\gamma_j = \Lambda_j \mu_j$, $\mu_j = \sqrt{4\pi^2 k^2 + q_j^2}$, $q_j = \sqrt{i\omega C_j/\Lambda_j}$, and $N$ is the total number of layers.

From $G_{up}$ and $G_{down}$, G can be calculated

$$G(k)^{-1} = G_{up}(k)^{-1} + G_{down}(k)^{-1} \quad (S16)$$

With G known, Eq. (S8) leads to $T_o$ (temperature change at air-water interface). The temperature profile in air and water can be written as below

$$\Delta T(k,z) = \Delta T_0(k,z)exp(\mu_1 z) \quad \text{for air } (z<0) \quad (S17)$$

$$\Delta T(k,z) = \Delta T_0(k,z)exp(-\mu_2 z) \quad \text{for water } (z>0) \quad (S18)$$

$$\Delta T_0(k) = G_0(k)P(k) \quad (S19)$$

The total deflection $\Delta\theta_t$ of the transmitted beam in water can be separated into three parts as illustrated Fig. S10.

$$\Delta\theta_t = \frac{n_{air}}{n_{water}} \times \frac{cos\theta_{air}}{cos\theta_{water}} \times \Delta\theta_1 + \Delta\theta_{inter} + \Delta\theta_2 \quad (S20)$$

$\Delta\theta_1$, $\Delta\theta_{inter}$, and $\Delta\theta_2$ are the beam deflection in the air phase, the air-water interface, and the water phase respectively. In air, the temperature field sensed by the probe beam is

$$\Delta T(r,z) = \int_0^\infty A_p exp(-\pi^2 k^2 w_p^2) \times exp(\mu_1 z) \times \Delta G_0(r,k)\, 2\pi k J_0(2\pi k r)dk \quad (S21)$$

where integration over $k$ is the inverse Henkel transform. The corresponding refractive index change is

$$\Delta n(r,z) = \left(\frac{dn}{dT}\right)_{air} \Delta T(r,z) \quad (S22)$$

Similarity, in the water phase

$$\Delta n(r,z) = \left(\frac{dn}{dT}\right)_{water} \Delta T(r,z) \tag{S23}$$

where $\left(\frac{dn}{dT}\right)_{air}$ and $\left(\frac{dn}{dT}\right)_{water}$ are thermo-optic coefficient of air ($9 \times 10^{-7}\ K^{-1}$) (35) and water ($-1.13 \times 10^{-4}\ K^{-1}$) (51), respectively. The integration over $k$ is the inverse Henkel transformation.

The differential beam deflection $d(\Delta\theta)$ caused by the local refractive index gradient is

$$nd(\Delta\theta) = \nabla_\perp \Delta n(r,z)\, ds \tag{S24}$$

where $\nabla_\perp \Delta n(r,z)$ is the gradient of the index of refraction perpendicular to s (the ray path), and $ds = \frac{dz}{\cos\theta}$ and $\nabla_\perp \Delta n(r,z) = \frac{d\Delta n}{dz}\sin\theta + \frac{d\Delta n}{dr}\cos\theta$. Hence, we have for the air and water side,

$$d\Delta\theta_1 = \frac{1}{n_{air}}\left(\frac{d\Delta n_{air}}{dz}\tan\theta_{air} + \frac{d\Delta n_{air}}{dr}\right)dz \tag{S25}$$

$$d\Delta\theta_2 = \frac{1}{n_{water}}\left(\frac{d\Delta n_{water}}{dz}\tan\theta_{water} + \frac{d\Delta n_{water}}{dr}\right)dz \tag{S26}$$

Next, we consider the beam angle change at the interface, $\Delta\theta_{inter}$ as shown Fig. S11, which can be caused by the refractive index change due to temperature and also by the surface deformation due to thermal expansion. For this purpose, since the thermo-optic coefficient of air is much smaller than that of water, we assume the refractive index of air as a constant of 1.

According to Snell's Law

$$n_{water} \times \sin(\theta_2) = n_{air} \times \sin(\theta_{air}) \tag{S27}$$

We put $\theta_1 = \theta_{air} + \Delta\theta_1$ and $n_{water} = n_{water} + \left(\frac{dn}{dT}\right)_{water} \times \Delta T_0$ into Eq. (S24),

$$\Delta\theta_{inter} = -\frac{1}{n_{water}}\tan(\theta_{water}) \times \left(\frac{dn}{dT}\right)_{water} \times \Delta T_0 + \left(\frac{n_{air}}{n_{water}} \times \frac{\cos\theta_{air}}{\cos\theta_{water}} - 1\right) \times \frac{dZ}{dr} \tag{S28}$$

where Z is the surface displacement (as a function of r). The first term arises from the refractive index change due to temperature change at the interface and the second term due to thermal expansion of the liquid. The second term (denote as $\Delta\theta_{inter-Z}$) in Eq. (S28) is obtained by considering the interface is rotate by $\frac{dZ}{dr}$ as shown in Fig. S12 due to the surface deformation. Black dash line represents the direction of the new interface at position $r$. The value factor of $\frac{dZ}{dr}$ is calculated by applying Snell law to the interface (black line)

$$n_{air}\sin\left(\theta_{air} + \frac{dZ}{dr}\right) = n_{water}\sin\left(\theta_{inter} + \Delta\theta_{inter-z} + \frac{dZ}{dr}\right) \quad (S29)$$

Considering $n_{air}\sin(\theta_{air}) = n_{water}\sin(\theta_{inter})$, and both $\frac{dZ}{dr}$ and $\Delta\theta_{inter-z}$ are small values. We get

$$\Delta\theta_{inter-z} = \left(\frac{n_{air}}{n_{water}} \times \frac{\cos\theta_{air}}{\cos\theta_{water}} - 1\right) \times \frac{dZ}{dr} \quad (S30)$$

We treat the surface deformation as if it is a solid, acknowledging in reality, the contribution of beam deflection of deformation at surface maybe smaller due to water fluidity. Under this assumption, the Hankel transform of surface displacement can be expressed as (*37,38*)

$$z(r,k) = \frac{1+\upsilon}{1-\upsilon}\alpha_T \frac{1}{\mu_2+\alpha} \frac{\beta^4 - 16\pi^4 k^4}{(\beta^2 - 4\pi^2 k^2)^2 - 16\alpha\beta\pi^2 k^2} T_0^{up}(r,k)$$

$$= \frac{1+\upsilon}{1-\upsilon}\alpha_T \frac{1}{\mu_2+\alpha} \frac{\beta^4 - 16\pi^4 k^4}{(\beta^2 - 4\pi^2 k^2)^2 - 16\alpha\beta\pi^2 k^2} G_0^{up}(r,k)P(r,k) \quad (S31)$$

$$\alpha = (-\omega^2/v_L^2 + 4\pi^2 k^2)^{1/2} \quad (S32)$$

$$\beta = (-\omega^2/v_T^2 + 4\pi^2 k^2)^{1/2} \quad (S33)$$

where $\upsilon$ is Poisson ratio of water, $v_L$ and $v_T$ are the longitudinal and transverse speed of sound of water, and $\alpha_T$ is the coefficient of thermal expansion (CTE) of water.

Integrating Eqs. (S25) & (S26) and adding Eq. (S28), we obtain the Now, we show the the total probe beam deflection $\Delta\theta_t$

$$\Delta\theta_t = \int_0^\infty A_p \exp(-\pi^2 k^2 w_p^2) - \frac{1}{n_{water}} \tan(\theta_{water}) \times \left(\frac{dn}{dT}\right)_{water} \times \Delta G_0(r,k)\, 2\pi k J_0(2\pi k \times D) dk + \left(\frac{n_{air}}{n_{water}} \times \frac{\cos\theta_{air}}{\cos\theta_{water}}\right) \int_0^\infty A_p \mu_1 \exp(-\pi^2 k^2 w_p^2) \int_{-\infty}^0 -\frac{1}{n_{air}} \tan(\theta_{air}) \times \left(\frac{dn}{dT}\right)_{air} \times$$

$$G_0(k)exp(\mu_1 z)\, 2\pi k J_0\left(2\pi k(D + z\times tan(\theta_{air}))\right) dz\, dk + \left(\frac{n_{air}}{n_{water}}\times\right.$$

$$\left.\frac{cos\theta_{air}}{cos\theta_{water}}\right)\int_0^\infty A_p exp(-\pi^2 k^2 w_p^2)\int_{-\infty}^0 -\frac{1}{n_{air}}\times\left(\frac{dn}{dT}\right)_{air}\times G_0(k)exp(\mu_1 z)\, 4\pi^2 k^2 J_1\left(2\pi k(D + \right.$$

$$\left. z\times tan(\theta_{air}))\right) dz\, dk + \int_0^\infty\int_0^\infty -A_p\mu_2 exp(-\pi^2 k^2 w_p^2)\times -\frac{1}{n_{water}}tan(\theta_{water})\times\left(\frac{dn}{dT}\right)_{water}\times$$

$$\Delta G_0(k)exp(-\mu_2 z) 2\pi k J_0\left(2\pi k(D + z\times tan(\theta_{water}))\right) dz\, dk + \int_0^\infty\int_0^\infty A_p exp(-\pi^2 k^2 w_p^2)\times$$

$$-\frac{1}{n_{water}}\times\left(\frac{dn}{dT}\right)_{water}\times \Delta G_0(k)exp(-\mu_2 z) 4\pi^2 k^2 J_1\left(2\pi k(D + z\times tan(\theta_{water}))\right) dz\, dk +$$

$$\left(\frac{n_{air}}{n_{water}}\times\frac{cos\theta_{air}}{cos\theta_{water}} - 1\right)\int_0^\infty exp(-\pi^2 k^2 w_p^2/2) z(r, K)\, 4\pi^2 k^2 J_1(2\pi k D)\, dk \tag{S34}$$

Note in this equation, the geometry equation $r = z\times tan(\theta_{water})$ is applied. The final beam deflection of beam exiting into air is

$$\Delta\theta = \frac{\Delta\theta_t}{\frac{n_{air}}{n_{water}}\times\frac{cos\theta_{air}}{cos\theta_{water}}} \tag{S35}$$

The real-space temperature distribution on water surface can be calculated by the following

$$T_0(r) = \int_0^\infty A_p exp(-\pi^2 k^2 \omega^2) G(k) 2\pi k J_0(2\pi k r)\, dk \tag{S36}$$

We use the $\sqrt{w_x w_y cos(\theta)}$ as $1/e^2$ radius of the beam size in the of Eq. (S34) and (S36), where $w_x$ and $w_y$ are the $1/e^2$ radius of the beam size in the two directions.

**SM5.3 Normalized beam deflection of SS-TBD**

We applied Snell's law to the transmitted beam (in air)

$$n_{equ}sin(\theta_t) = n_{air}sin(\theta_{air}) \tag{S37}$$

where $\theta_t = \theta_{air} + \Delta\theta$ and $n_{equ} = n_{air} + \Delta n_{equ}$. $\Delta\theta$ is the measured beam deflection (on the order of μrad), which is much smaller than $\theta_{air}$. $\Delta n_{equ}$ is the equivalent refractive index change of air between sample and detector due to the photomolecular effect. $\Delta n_{equ}$ is much smaller than $n_{equ}$ (equivalent refractive index). Neglecting higher order terms,

$$\Delta n_{equ} = \frac{\Delta\theta}{tan(\theta_{air})} \tag{S38}$$

As a result, we define $\frac{\Delta\theta}{tan(\theta_{air}) \times I}$ as the normalized beam deflection.

## SM6. Measurement of Transmission of Vapor Above Air-water Interface

The optical path of the vapor phase transmission measurement is shown in Fig. S18. We added the green laser into the sample compartment of a commercial Cary Series UV-vis-NIR spectrometer.

A high power 532 nm CW laser (1.5 W) was used as the pump laser. Two lenses with different focal length (20 mm and 60 mm) were utilized to expand the laser and the laser covers the air-water interface in the glass container (diameter 30 mm). Two mirrors were utilized to change the incident angle of pump. The polarization of pump was changed by a half-wave plate. The limited space of the sample compartment does not allow complete expansion and collimation of the laser beam, and hence the pump laser has some divergence. As a result, the incident angle is in the range of 40° to 50° (average is 45°), the polarization is a TM-polarization, and the power and $1/e^2$ radius of the expanded high power green laser close to air-water interface are 1.4 W and ~9 mm, respectively. We have checked that the laser beam does not scatter light into the detector.

A CaF lens (focal length 20 mm) with high IR transmission was used to focus the light source from the spectrometer to a beam diameter ~1 mm at the focal point, without much loss of the power. The air-water interface is adjusted and placed at ~4 mm below the height of beam at focal plane.

The transmittance measurement consists of the following steps. First, all accessories are placed inside the chamber and dry air is used to purge the instrument. The transmission spectrum is measured and used as the reference. Then, water is added to the container and heated to 53 ºC. The transmission spectrum is measured again and normalized to the dry-air spectrum to obtain the transmittance (T). After the green laser is turned on, transmission spectrum is measured again. Each spectrum takes ~2 minutes.

## SM7. Raman Spectrum

We used a commercial confocal Raman spectrometer (HORIBA Scientific) to measure the spectrum of liquid water and vapor by varying the position of focal plane in four different configurations (Fig. S19): the laser focal point is above (AS), on (OS), below (BS) the water surface, or with the emersion lens (EL) configuration. The EL configuration avoids surface effect and minimizes any potential heating effect, and hence can be considered to measure Raman spectrum of the bulk water. The confocal configuration of the Raman system ensures most signals come around the focal point in the vapor phase in the AS configuration and in the water in the BS configuration. However, due to much higher density of water, the AS configuration unavoidable also has some signals from water even in the confocal configuration and the signal is mostly from water in AS geometry when focal point is not too far away from surface (Fig. S33b).

The wavelength of the laser in Raman system is 532 nm. The $1/e^2$ radius of the laser is 1.1 mm at the back focal plane of the lens. The working distances of 50X and 100X focal lens are 26.5 and 0.21 mm, respectively. The beam $1/e^2$ beam radius at the focal plane of 50X and 100X focal are 1.0 and 0.5 μm, respectively. The maximum power of laser at the focal plane of 50X and 100X lens are 33 mW and 17 mW, respectively.

In the measurement of the liquid phase (Fig. S33, red curve in Fig. 3d, and red curve in Fig. 3e) the time to acquire Raman spectrum is 30 s. In the measurement of vapor phase using 100X lens (black curve in Fig. 3e and Fig. S20), the time to acquire Raman is 1200 s. Intensity of liquid phase Raman scattering is ~1500 times of the vapor phase in our measurement. In the measurement of vapor phase above heated water surface using the 50X lens (black curve in Fig. 3d), the time to acquire Raman spectrum is 300s.

In the AS configuration, the divergent laser beam after the focal point meets the water surface with a range of angles (0-25° for 50X and 0-60° for 100X) and has both TE and TM polarizations, generating water clusters. Since the 50X lens has a long working distance, water can be heated to carry clusters up, and the focal point can be placed a few mm above the water surface. For 100X lens, the short work distance does not allow us to use heated sample, since the water interface recesses during data collection.

Due to different magnitudes of Raman signals between OS and AS, we normalize the signals at the maximum signal in the range of 2900-3900 cm$^{-1}$ for direct comparison purposes.

As illustrated in Fig. S33a, the Raman laser can also heats up water surface in the AS configuration similar to the experiments in Fig. 1. In the AS configuration with focal plane only 30 μm above water surface, Raman signals (Fig. S33b) are exclusively from the water surface. The lower intensity below 3425 cm$^{-1}$ than that of the EL configuration indicates lower ratio of OH-stretching oscillators to nonhydrogen-bonded OH-stretching oscillators, which happens when the temperature of water rises as explained in a previous report (*52*). The temperature increase is consistent with the temperature rise of water surface due to photomolecular effect measured by IR camera and thermocouple (Figs. 1b-d).

## SM8. Measurement of 2ω Raman Scattering

The optical path for the 2ω Raman scattering experiment is shown in Fig. S21. A high power 532 nm CW green laser (work as pump) was expanded by two focal lens and modulated by a mechanical chopper (modulation frequency f=200 Hz) placed at the focal plane of the two objective lenses to get closer to the ideal square wave modulation. The power and $1/e^2$ radius of the expanded high power green laser are 0.7 W and 2.8 mm, respectively. Two mirrors were utilized to change the incident angle of pump. The polarization of the pump laser was changed by a half-wave plate. The mechanical chopper provides a reference signal (f) to lock-in amplifier. A 550 nm long pass filter and a 600 nm long path filter were placed between the sample and a photodiode detector to filter the elastic scattering of the green laser. The photodiode was placed 1 cm away from the beam spot on the surface and not directly exposed to the incident and the reflected laser. The photodiode was connected to a lock-in amplifier. The harmonic number of lock-in amplifier is set as 2. Time constant of the lock-in amplifier is 100 s. In the measurement, we vary the power of the laser without changing any of the set-up since the amplitude of the signal is sensitive to the position of the detector.

## SM9. Mass Measurement Using a Constrained Beam

**SM9.1 Measurement**

The optical path for such measurement is shown in Fig. S23. A high power 532 nm CW laser (1.5 W) was expanded by two objective lenses with focal length of 20 mm and 150 mm to cover the air-water interface in the glass container (diameter 30 mm). Two mirrors were utilized to change the incident angle of the pump laser. The polarization of the pump laser was changed by a half-wave plate. The pump laser was modulated by a mechanical chopper, which provides a reference signal to the lock-in amplifier.

A pen laser (635 nm) was used as the probe, which is perpendicular to the plane of incidence of the pump laser. A 600 nm long pass filter was placed before a position sensing detector to avoid the influence of the scattered pump laser on the measurement of the beam position shift, although we found no difference in signals with and without the filter because the directions of the pump and probe are orthogonal. The position sensitive detector was connected to a lock-in amplifier to measure the deflection of the modulated beam. Time constant of the lock-in amplifier is 30 s for f>100 Hz and 100 s for f<100 Hz.

The length (L), width (W), and thickness (D) of PMMA beam are 150 mm, 30 mm, and 0.25 mm. These values are chosen with consideration of the amplitude of signal, the noise, and the modulation frequency. A water container was put on the top center of the PMMA beam. A mirror was stuck under the bottom and 30 mm away from center of the cantilever, which was not exposed to the pump beam. Two ends of the beam is glued to metal holder.

The modulated signal (AC signal) should not be affected by the natural evaporation (DC signal).

## SM9.2 Model

Assuming the forced is applied in the middle of the beam, the displacement in the middle of the bending beam is (*37*)

$$\Delta z = FL^3/(3EM) \tag{S39}$$

where L, w, and h are the length, width and height of bending cantilever, F is the force applied in the middle of the beam, E is the elastic constant of materials, M is moment of inertia following

$$M = \frac{1}{768}wh^3 \tag{S40}$$

Resonance vibration frequency $\omega_r$ is calculated by

$$\omega_r = \sqrt{\frac{K}{m}} \tag{S41}$$

where m is the total mass of the bending cantilever, glass container, mirror and water, $K = (3EM)/L^3$.

For forced vibration with amplitude $F_0$ and angular frequency $\omega$, amplitude of the vibration $Z_0$ is (*53*)

$$Z_0 = \frac{F_0}{K} \cdot \frac{1}{\left(\left(1-\frac{\omega^2}{\omega_r^2}\right)^2 + \left(2\delta\frac{\omega}{\omega_r}\right)^2\right)^{1/2}} \tag{S42}$$

$\delta$ is the damping coefficient.

The length (L), width (W), and thickness (D) of PMMA are 150 mm, 30 mm, and 0.25 mm, respectively. The total mass of glass container, mirror, and water are ~30g.

The beam deflection signal is equal to $\frac{2Z_0}{L/2} = \frac{4Z_0}{L}$ (Fig. 4a)

In actual measurements, the force is applied via laser-induced weight change in the area of the glass container instead of middle of the beam. We use the signal of detector reading caused by the mass change during natural evaporation (without light) as a calibration (1 g/V) to infer the mass change created by the modulated laser beam.

## SM10. Absorptance Estimation

We estimate the effective convective heat transfer coefficient $h$ under nature evaporation of water in glass container using the following equation:

$$h(T_a - T_s)(2A_{top-outer} + A_{side}) = \dot{m}LA_{top-inner} \tag{S43}$$

where $T_a$ is the ambient temperature (22.8 °C), $T_s$ is the areal-averaged surface temperature of water under nature evaporation (20.54 °C) measured by IR camera, $A_{top-inner}$ is the top inner surface area of the glass container (7.07×10$^{-4}$ m$^2$), $A_{top-outer}$ is the top outer surface area of the glass container (8.55×10$^{-4}$ m$^2$), $A_{side}$ is the lateral surface area of the container (1.04×10$^{-3}$ m$^2$), $\dot{m}$ is the measured nature evaporation rate (0.09 kg/m$^2$-h), and $L$ is the laten heat of water (2.264×10$^6$ J/kg). We can get $h \approx 7.96$ W/m$^2$-K.

Optical path for the absorptance estimation for open system (without cover) and sealed system (with PS cover) is shown in Fig. S24. IR camera (FLIR C5) is used to measure the areal average temperature of the container surface.

As shown in Fig. S25, open system (without PS cover) under power ($P$) of 1.4 W of TM-polarized green laser illumination with incident angle $\theta = 45°$, the areal average temperature of PS container $\bar{T}_{TM}$ is 21.08 °C while under TE-polarized green light illumination $\bar{T}_{TE}$ is 20.63 °C. Without green laser illumination, the average temperature of PS container $\bar{T}$ is 20.54°C. Assuming the same heat transfer coefficient $h$, the absorptance ($\alpha$) follows.

$$P\alpha = h(\bar{T}_{TM} - \bar{T})(2A_{top-outer} + A_{side}) \tag{S44}$$

As a result, $\alpha$ is estimated to be 0.84%.

In the above estimation of open system, we neglected the fact that heat transfer to water might change since air temperature is slightly cooler due to cluster dissociation (Fig. 2a). We also conducted an experiment that covers the container with PS cover (sealed system) under green laser illumination. In this case, clusters recondense, and the photomolecular effect leads to the water heating above the ambient. Assuming the same effective heat transfer coefficient, the absorptance obtained does not differ much from inferred above.

## SM11. Summary of the Measurements

We summarize the 13 performed experiments to probe the photomolecular effect in Table S6 to show what each experiment interrogates. For each observation (polarization, wavelength, incident angle, molecular clusters, and temperature response), we have at least two measurements to corroborate with each other. The table shows that the polarization effect showed up in 8 different experiments.

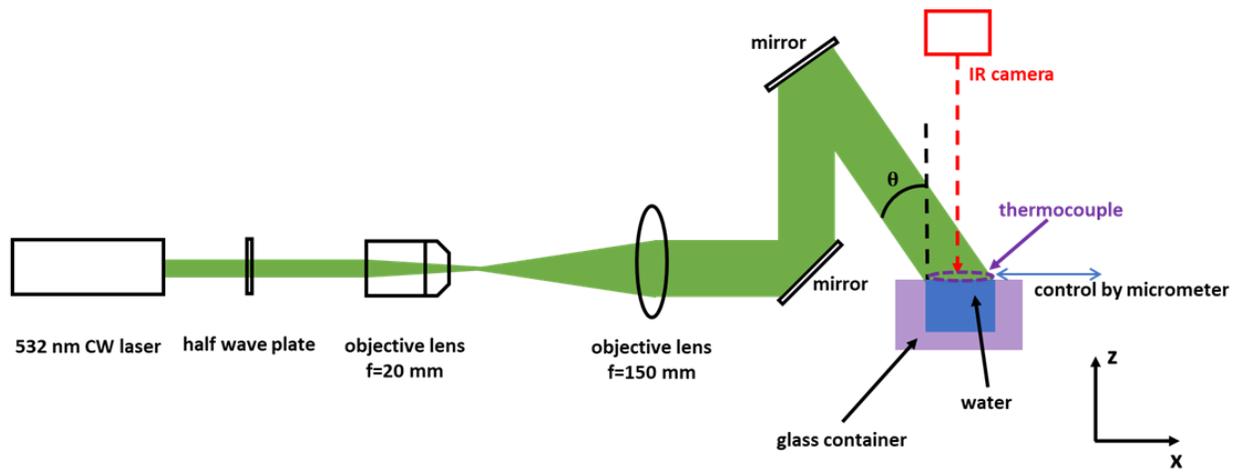

**Fig. S1. Optical path for water surface temperature measurement.**

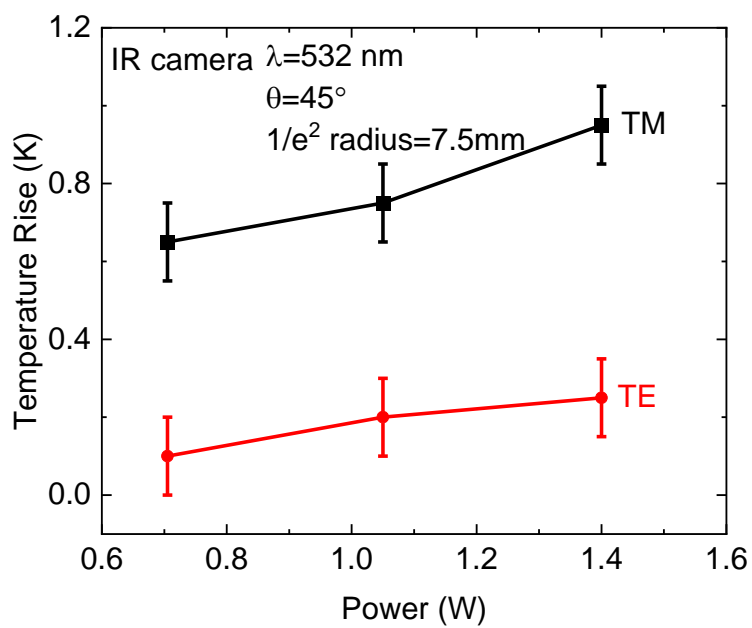

**Fig. S2. Water surface temperature rise as a function of laser power**. The radius of air-water interface is 15 mm. Incident laser power is 1.4 W.

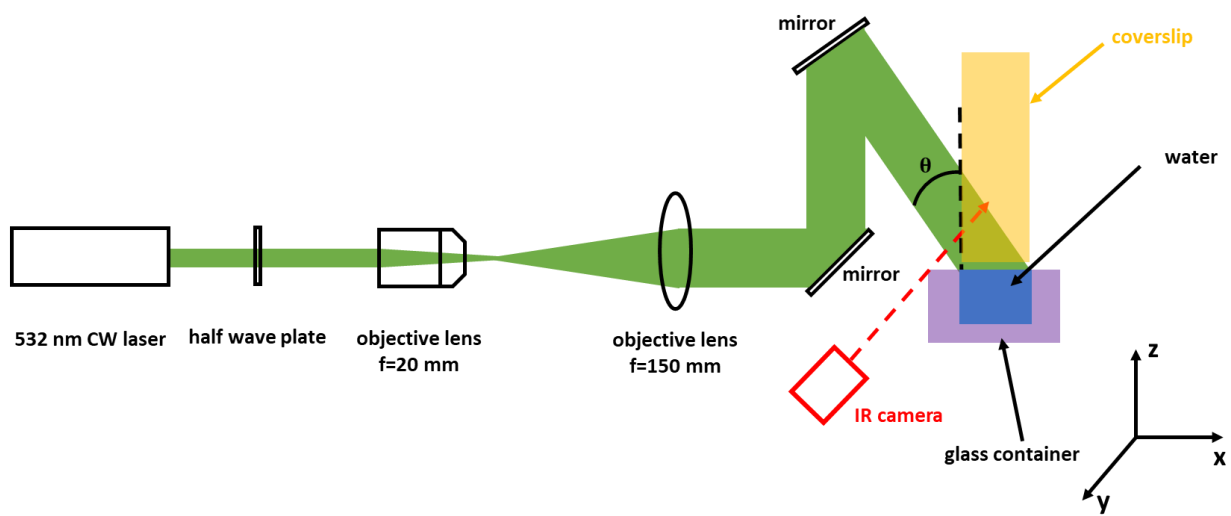

**Fig. S3. Optical path of vapor phase temperature distribution image above water surface.**

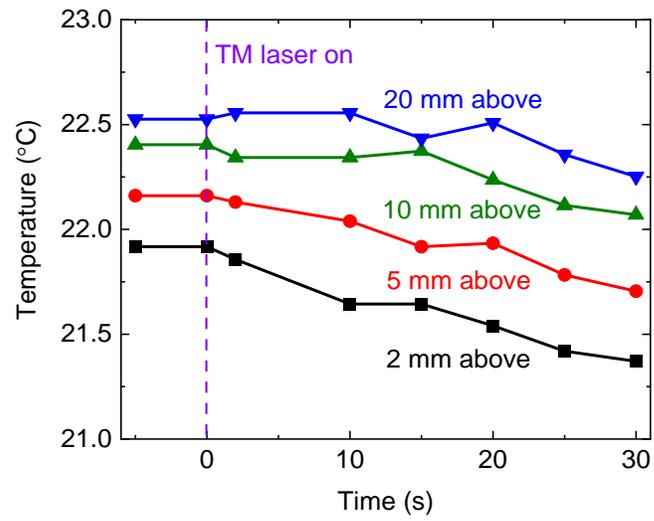

**Fig. S4. Temperature response of vapor phase temperature at different height above air-water interface.** Time zero is defined when the laser turns on. The incident angle of light is 45˚.

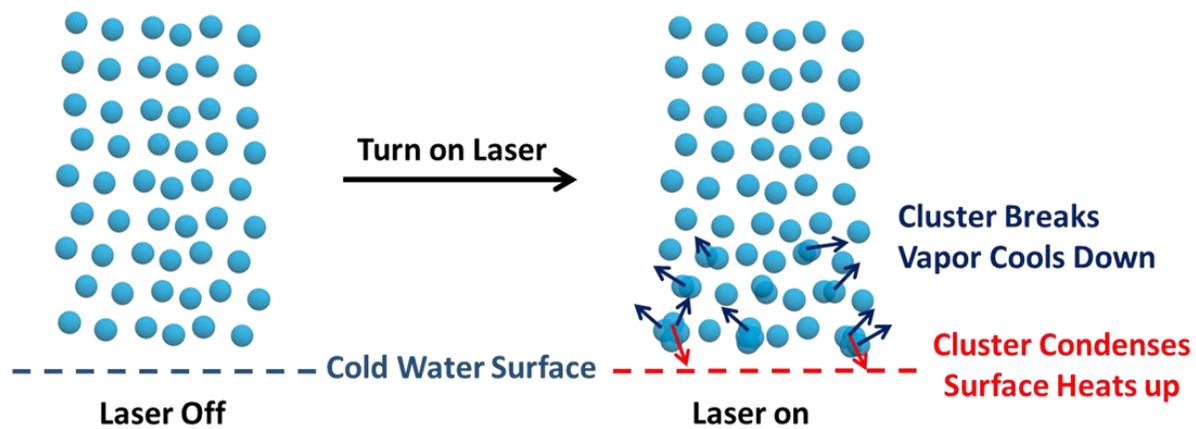

**Fig. S5. Schematic of temperature response in the vapor phase and on liquid surface due to the photomolecular effect.** Some clusters condense onto the water surface, heating up water. Some clusters break up in air, cooling down air.

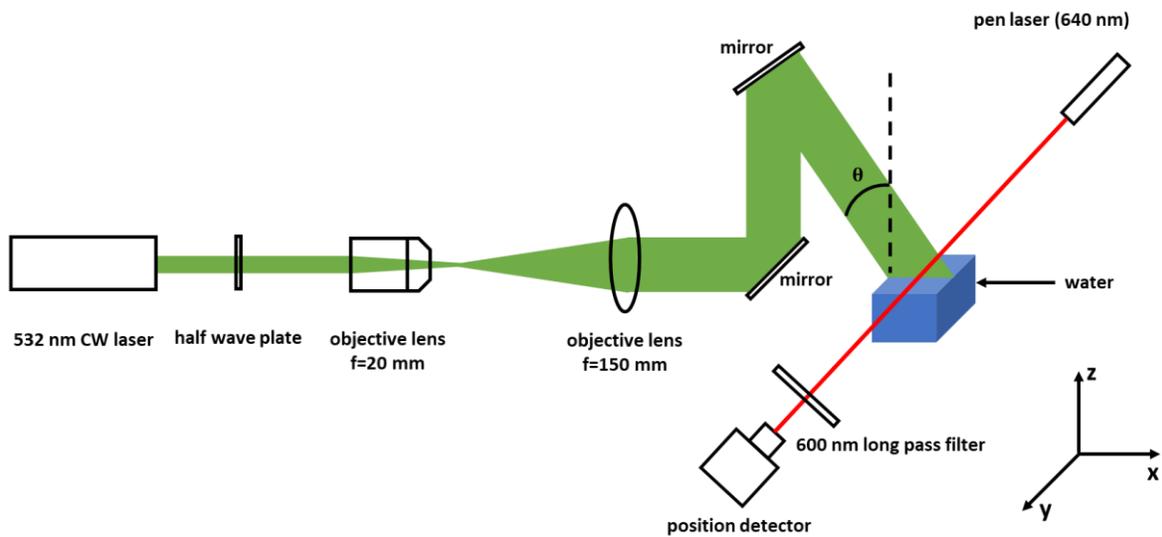

**Fig. S6. Optical path of the mirage effect of vapor phase above air-water interface.**

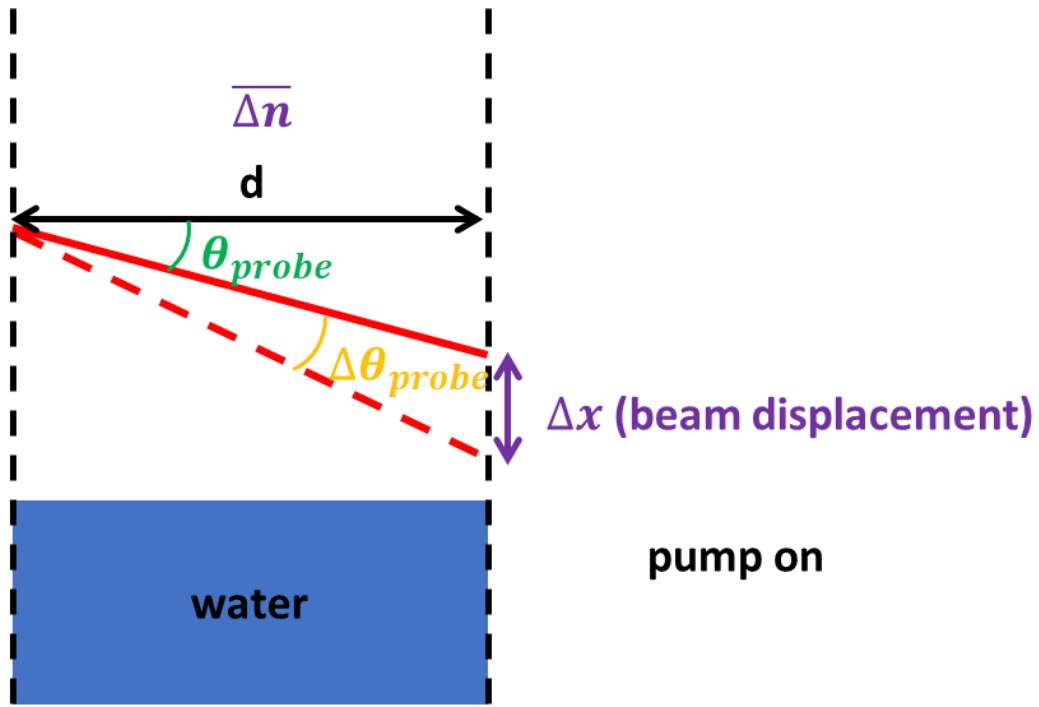

**Fig. S7. Geometry of the measurement of Mirage effect.** Red solid line represents the path of probe light without pump. Red dash line represents the path of probe light with pump on.

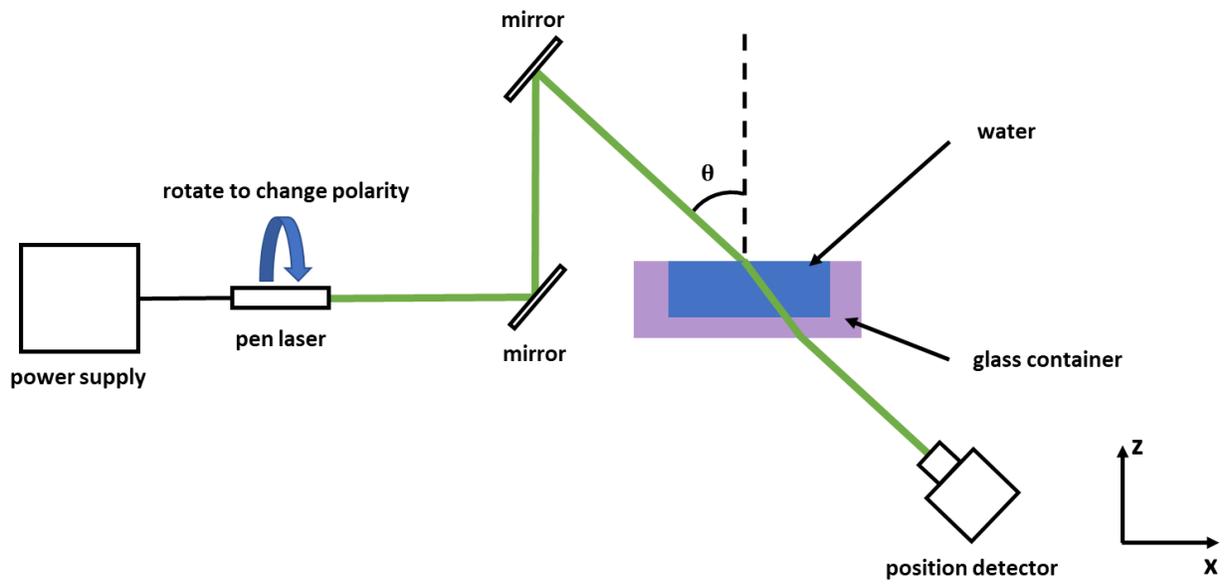

**Fig. S8. Optical path of steady-state transmitted beam deflection (SS-TBD) measurement.**

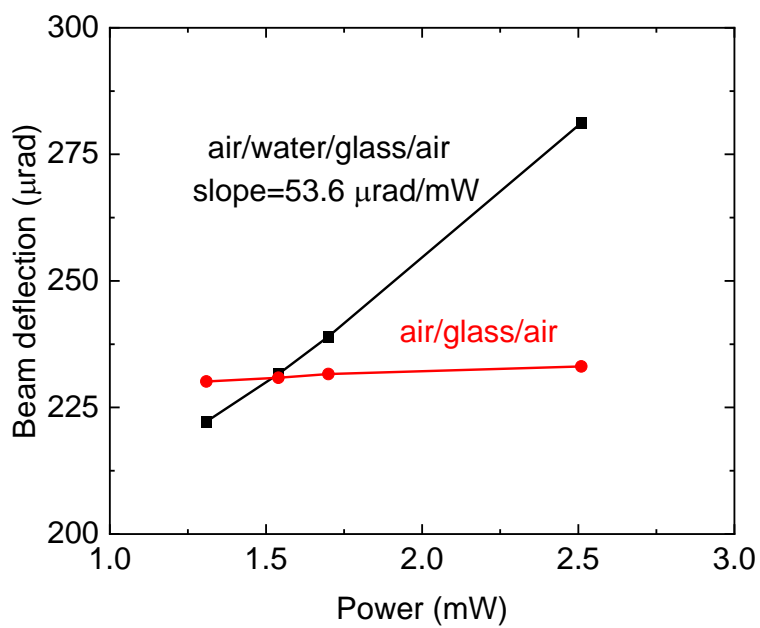

**Fig. S9. Beam deflection signal of SS-TBD as a function of power.** The wavelength of the laser is 635 nm. The polarization of the laser is TM. Black squares represent air-water/glass/air sample. Red circles represent air/glass/air sample.

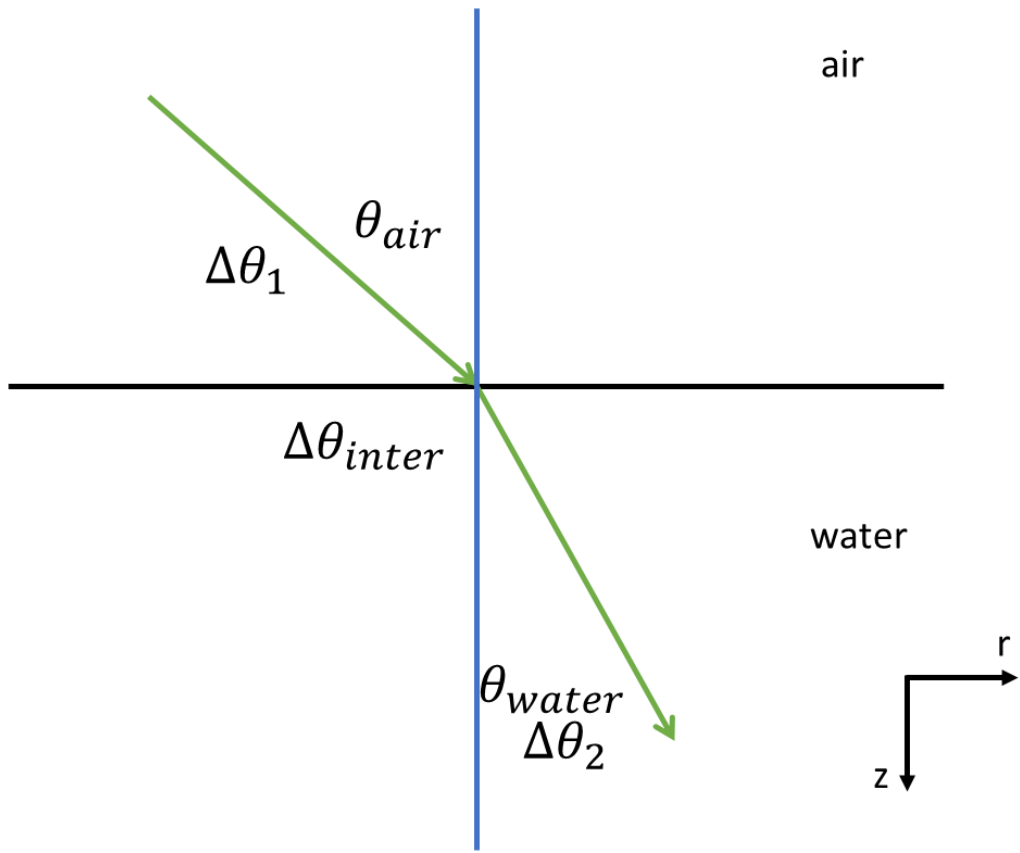

**Fig. S10. Display of beam deflections from temperature field at air-water interface.** $\Delta\theta_1$, $\Delta\theta_{inter}$, and $\Delta\theta_2$ are the beam deflection in the air phase, air-water interface and water phase respectively.

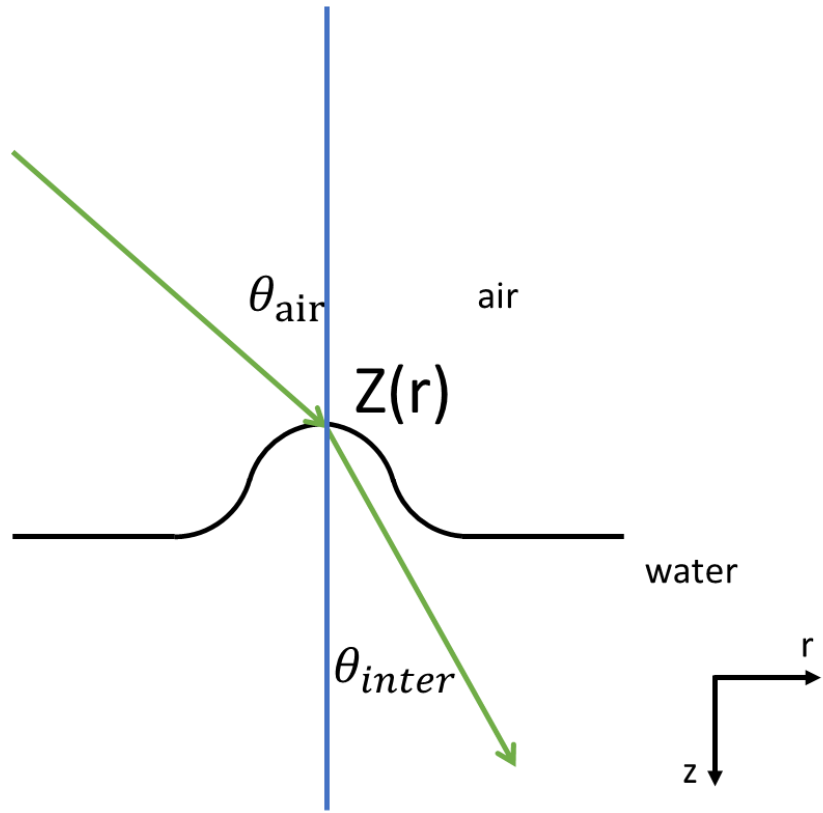

**Fig. S11. Display of beam deflection from air-water interface displacement.** The origin of the coordinate is the intersection of light and air-water interface.

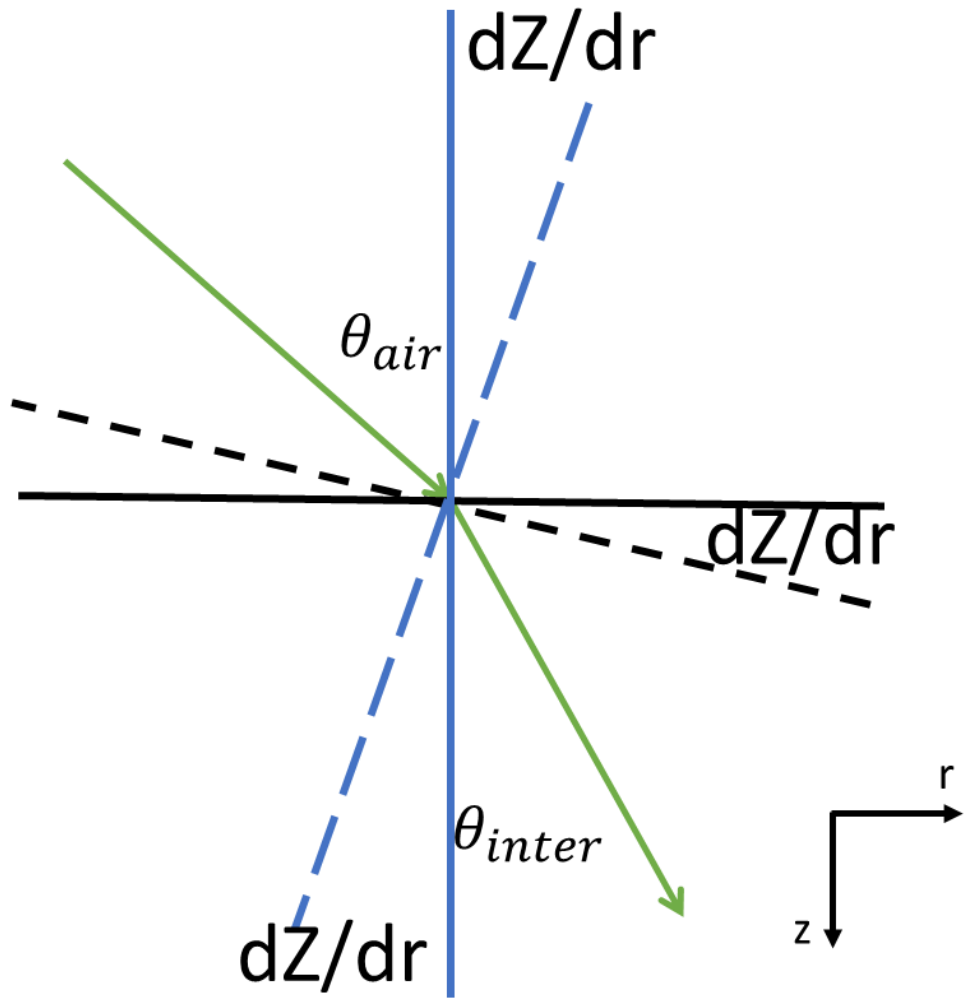

**Fig. S12. Display and labels of air-water interface displacement.**

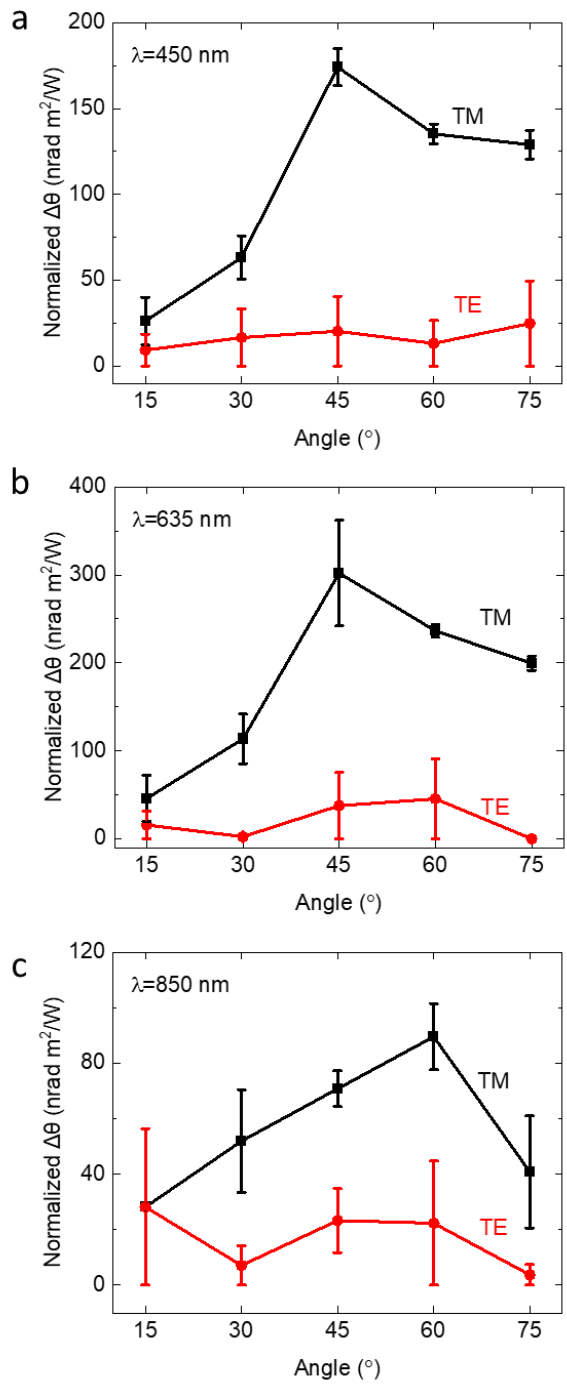

**Fig. S13. Normalized beam deflection as a function of the incident angle.**

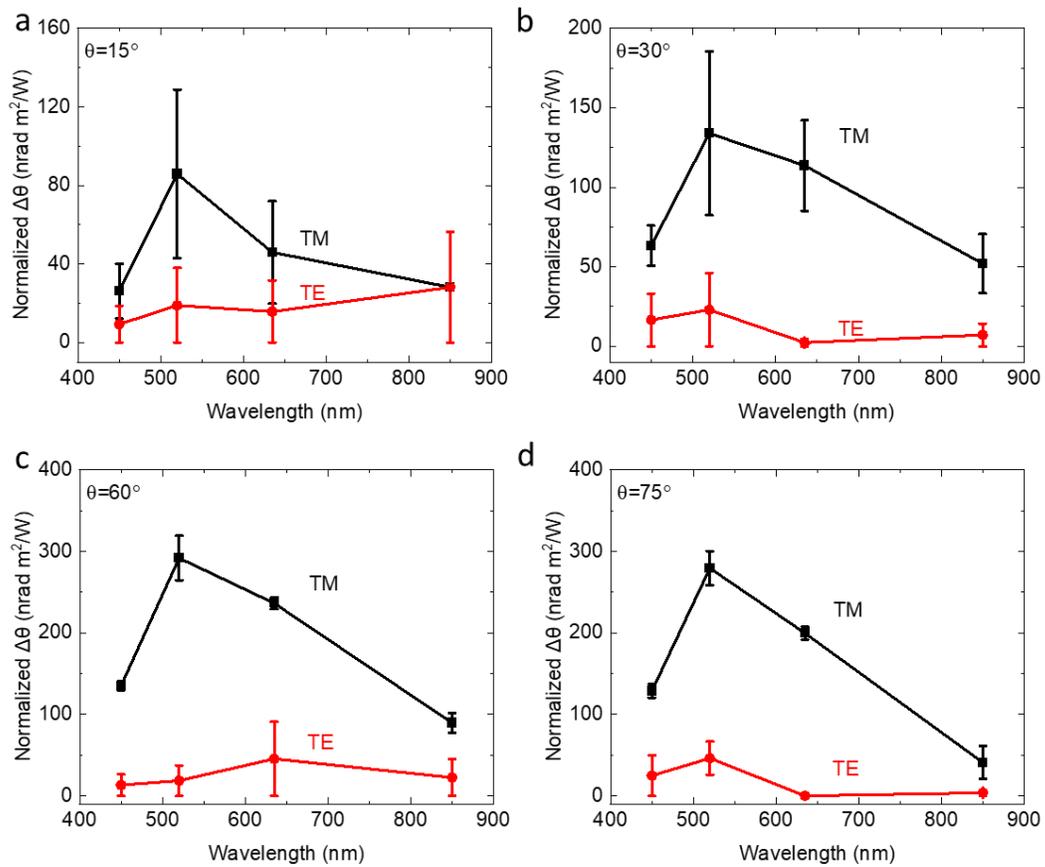

**Fig. S14. Normalized beam deflection as a function of wavelength of pen lasers.** (a) The incident angle = 15˚. (b) The incident angle = 30˚. (c) The incident angle = 60˚. (d) The incident angle = 75˚. Black squares represent TM laser. Red circles represent TE laser.

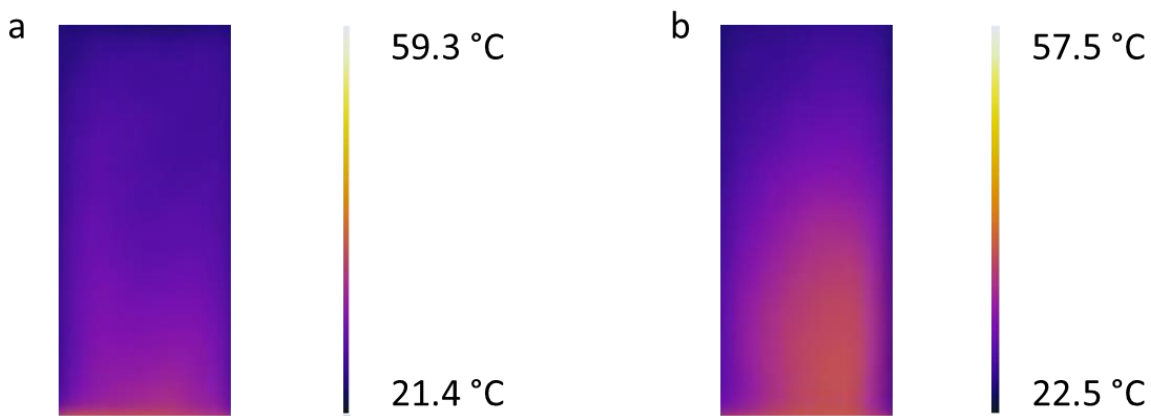

**Fig. S15. IR image of the coverslip above hot water surface.** (a) Under TM-polarized laser illumination (wavelength 532 nm, power 1.4 W, incident angle 45°, $1/e^2$ radius 7.5 mm). The radius of air-water interface is 15 mm. (b) Without laser illumination. Fig. 2g&h plots the temperature distribution along the middle of the coverslip.

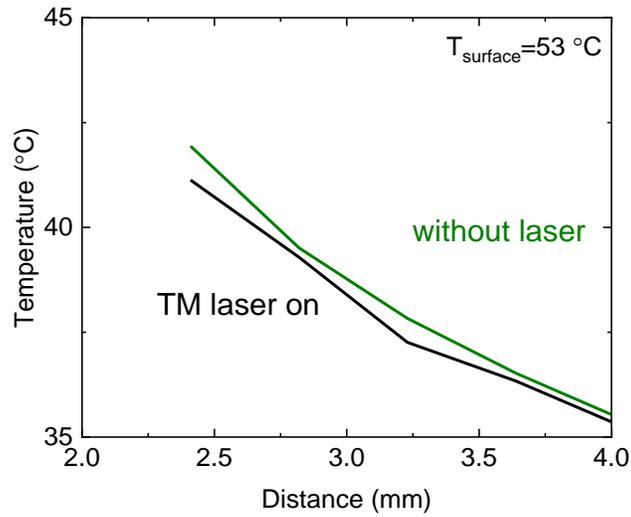

**Fig. S16. Vapor phase temperature distribution above hot water.** Black solid line represents steady-state vapor temperature profile with TM-polarized light (wavelength 532 nm, power 1.4 W, incident angle 45°, $1/e^2$ radius 7.5 mm). Green solid line represents steady-state vapor temperature profile without laser. The radius of air-water interface is 15 mm. Surface temperature of water is 53 °C before laser turns on. The polarization of the laser is TM.

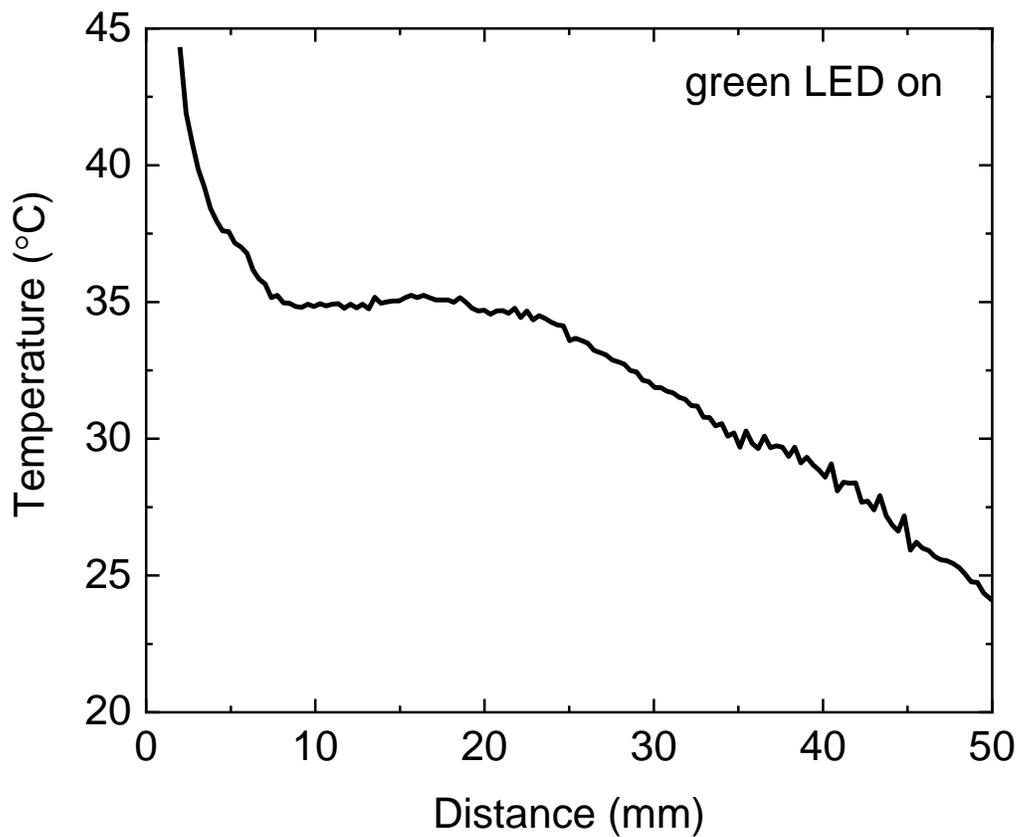

**Fig. S17. Vapor phase temperature distribution under LED illumination.** The wavelength of LED is 520 nm. The incident angle is 30˚. The surface temperature is ~55 ˚C before LED on. The intensity of green LED is 1000 W/m$^2$.

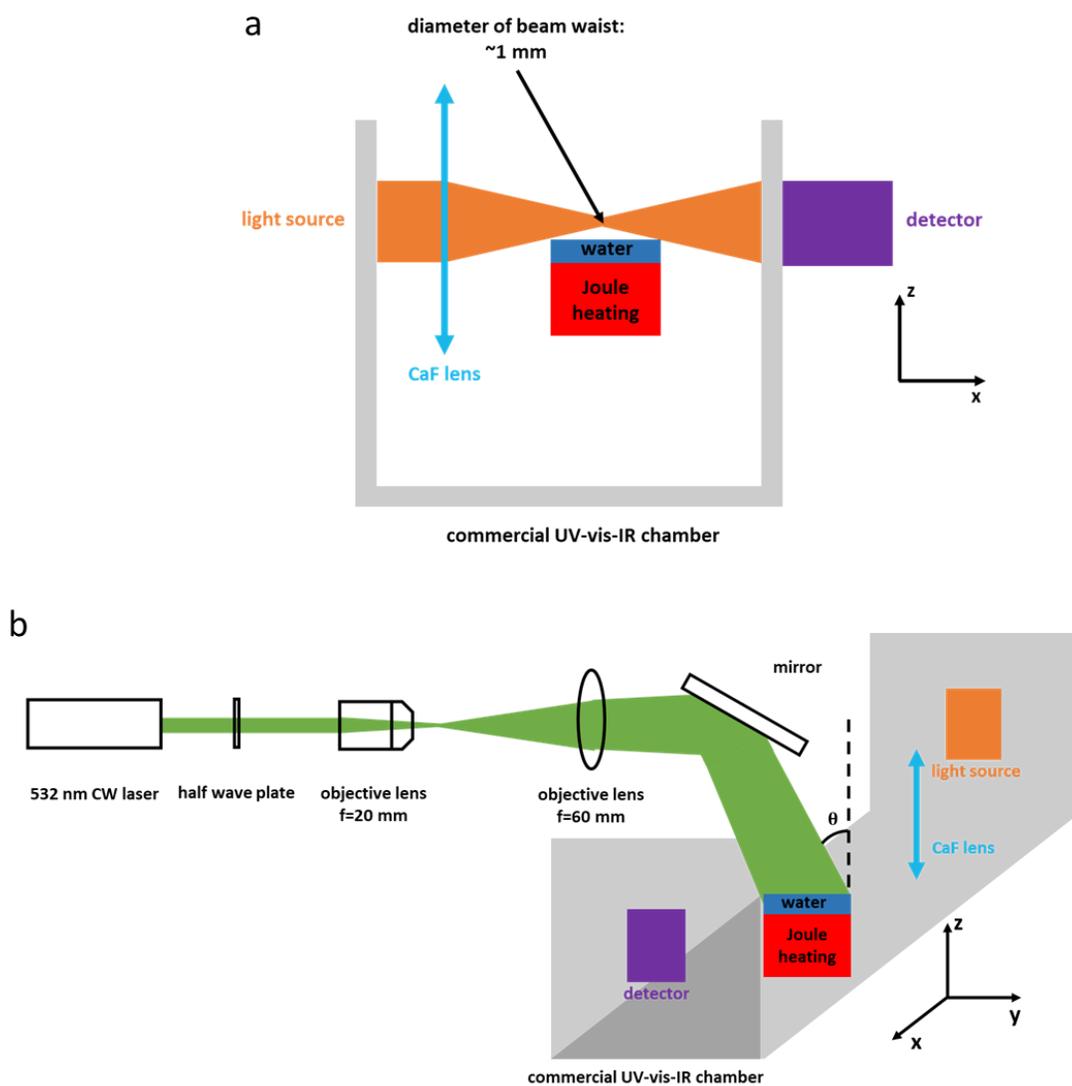

**Fig. S18. Optical path of vapor phase absorption measurement.** Grey represents the commercial UV-vis-IR chamber. (a) Front view of set-up. (b) Side view of set-up.

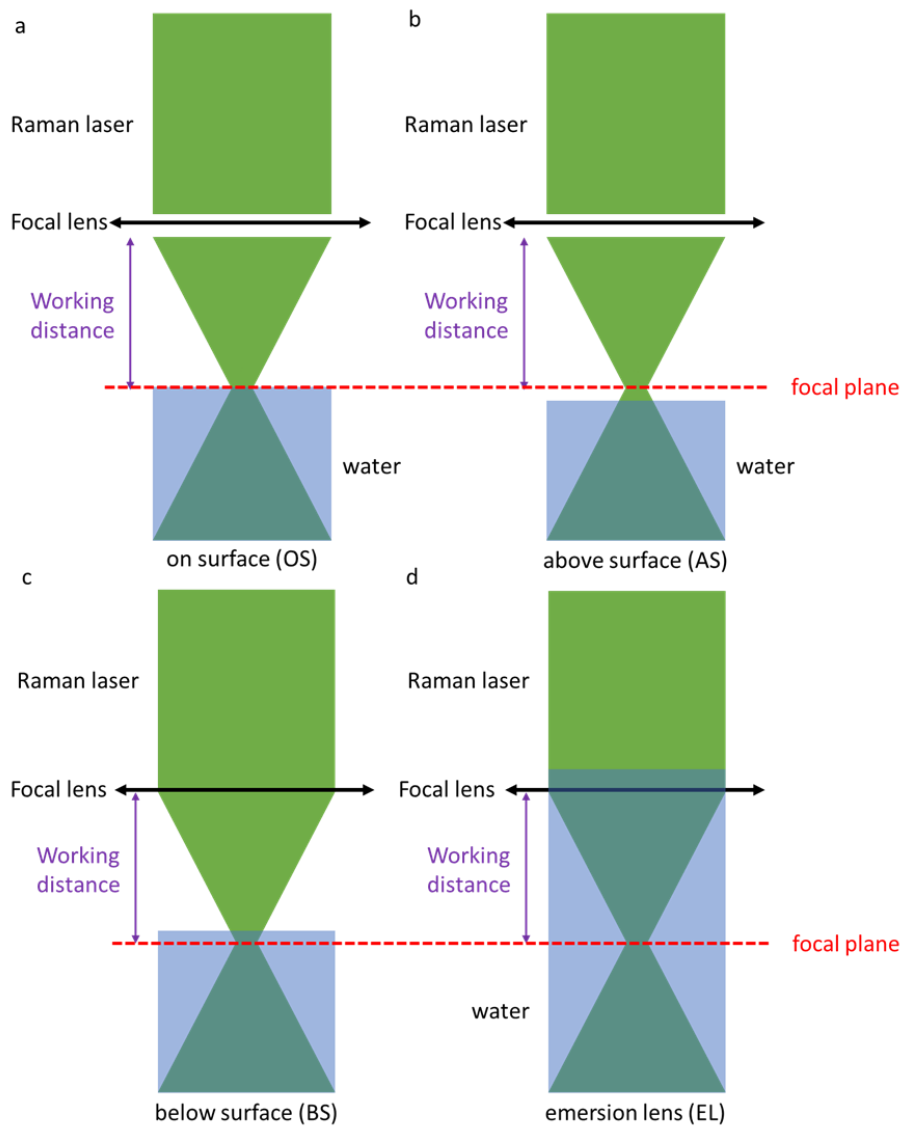

**Fig. S19. Schematics of Raman measurement.** The focal plane is at position of (a) on surface (OS), (b) above surface (AS), (c) below surface (BS), and (d) emersion lens (EL).

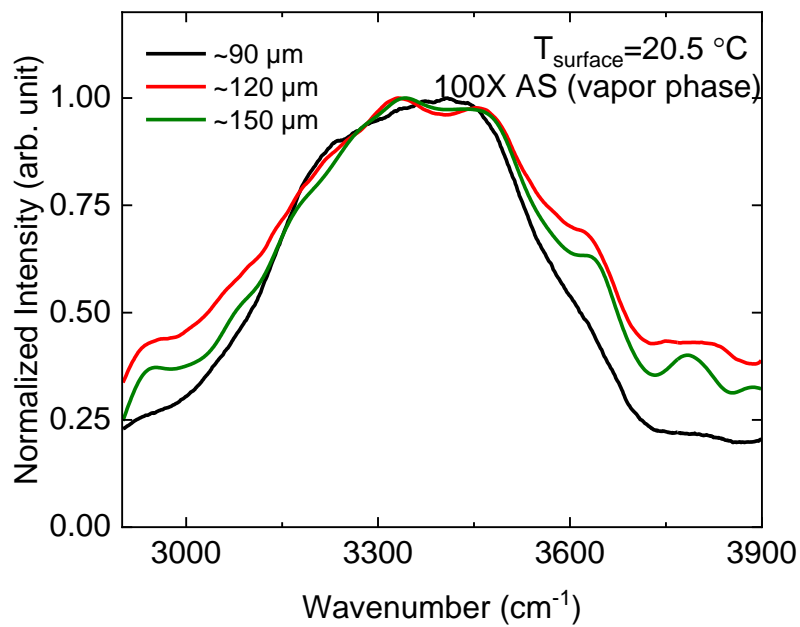

**Fig. S20. Evolution of Raman spectra.** The working distance of 100X lens is 0.21 mm. The $1/e^2$ radius of green laser (532 nm) at beam waist is ~0.5 μm. The average distance of beam waist above air-water interface is ~90 μm (black line), 120 μm (red line), and 150 μm (green line) during each measurement.

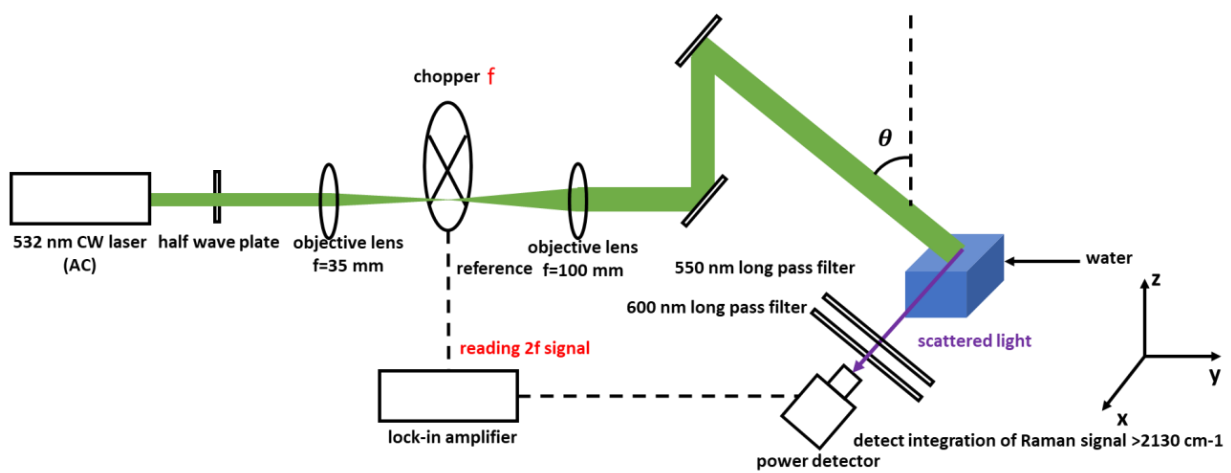

**Fig. S21. Optical path for 2𝜔 Raman measurement.**

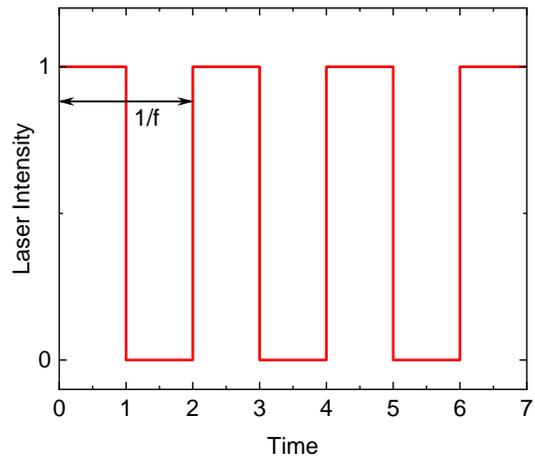

**Fig. S22. Waveform of ideal square wave.**

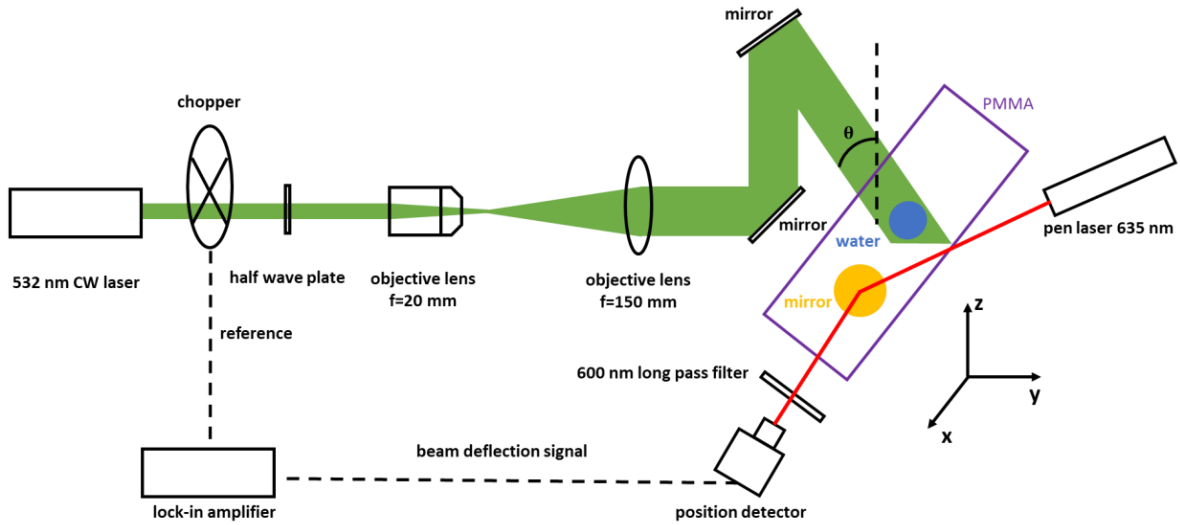

**Fig. S23. Optical path of modulated cantilever beam measurement.** The glass container of the water is not displayed in the figure.

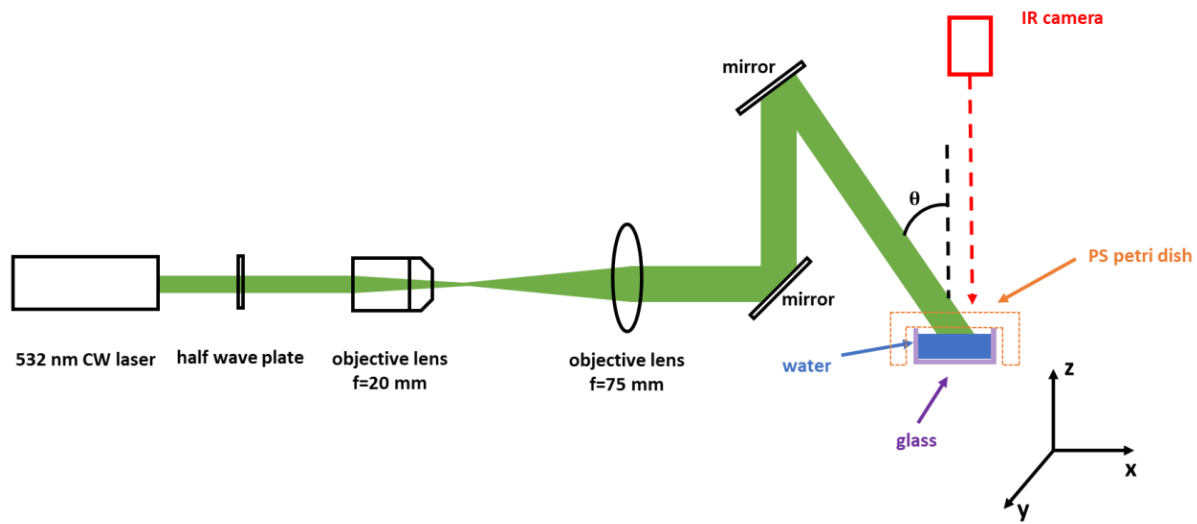

**Fig. S24.** Optical path for the absorption estimation measurement.

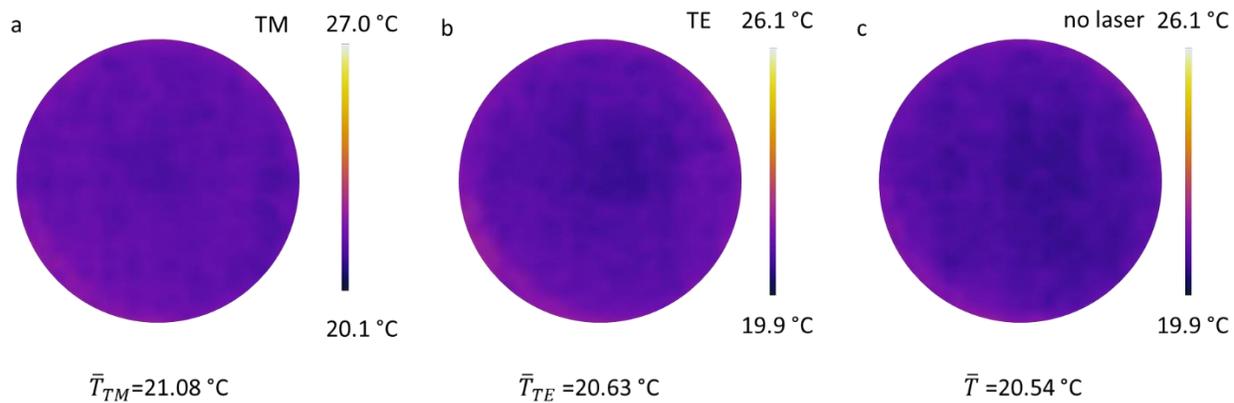

**Fig. S25. IR image the absorption estimation measurement**. (a) IR image under TM-polarized illumination (wavelength 532 nm, power 1.4 W, incident angle 45°, $1/e^2$ radius 3.75 mm) (b) IR image under TE-polarized illumination. (c) IR image without laser.

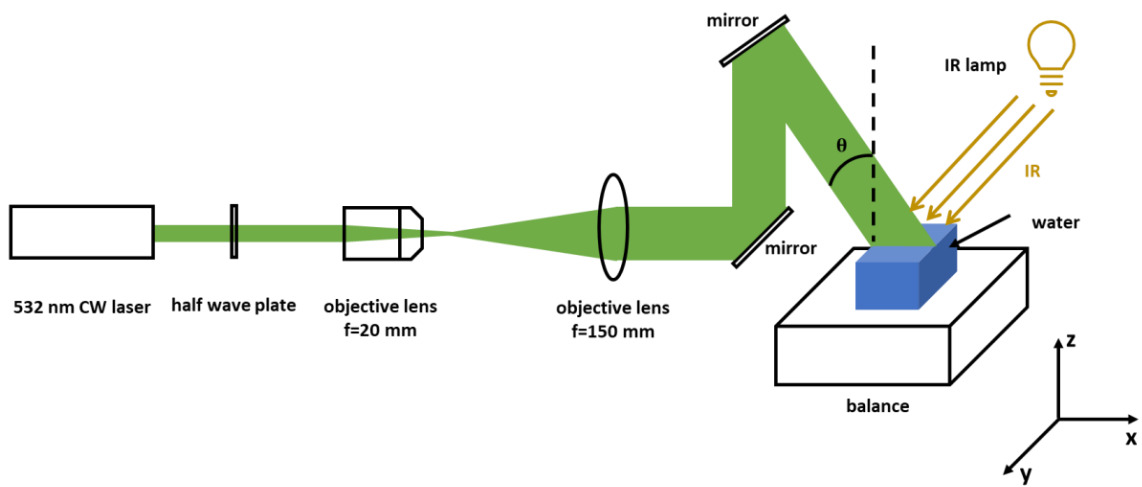

**Fig. S26. Optical path of evaporation measurement with IR lamp.** Glass container of the water is not displayed in the figure.

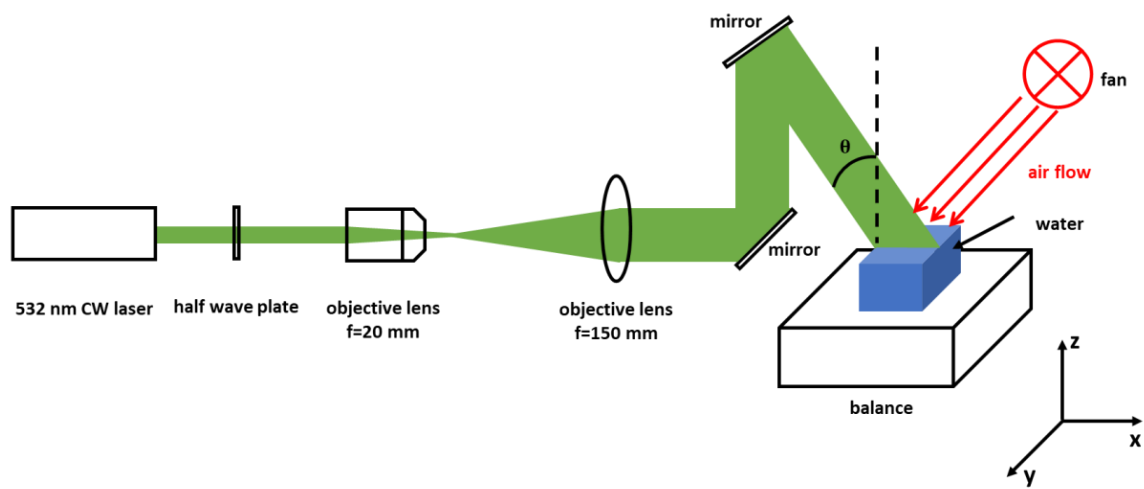

**Fig. S27. Optical path of evaporation measurement with fan.** Glass container of the water is not displayed in the figure.

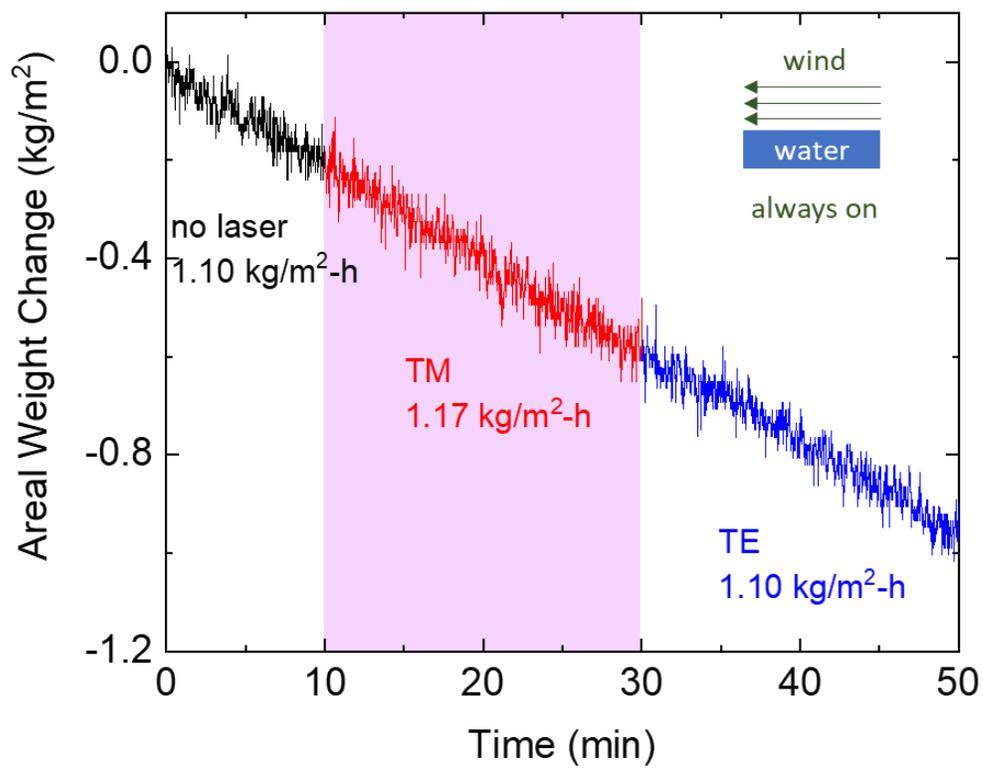

**Fig. S28. Areal weight changes as a function of water under air flow.** The power of laser is 1.0 W and the radius of air-water interface is 15 mm. Power/areal of water surface ≈ 1400 W/m² (1.4 suns).

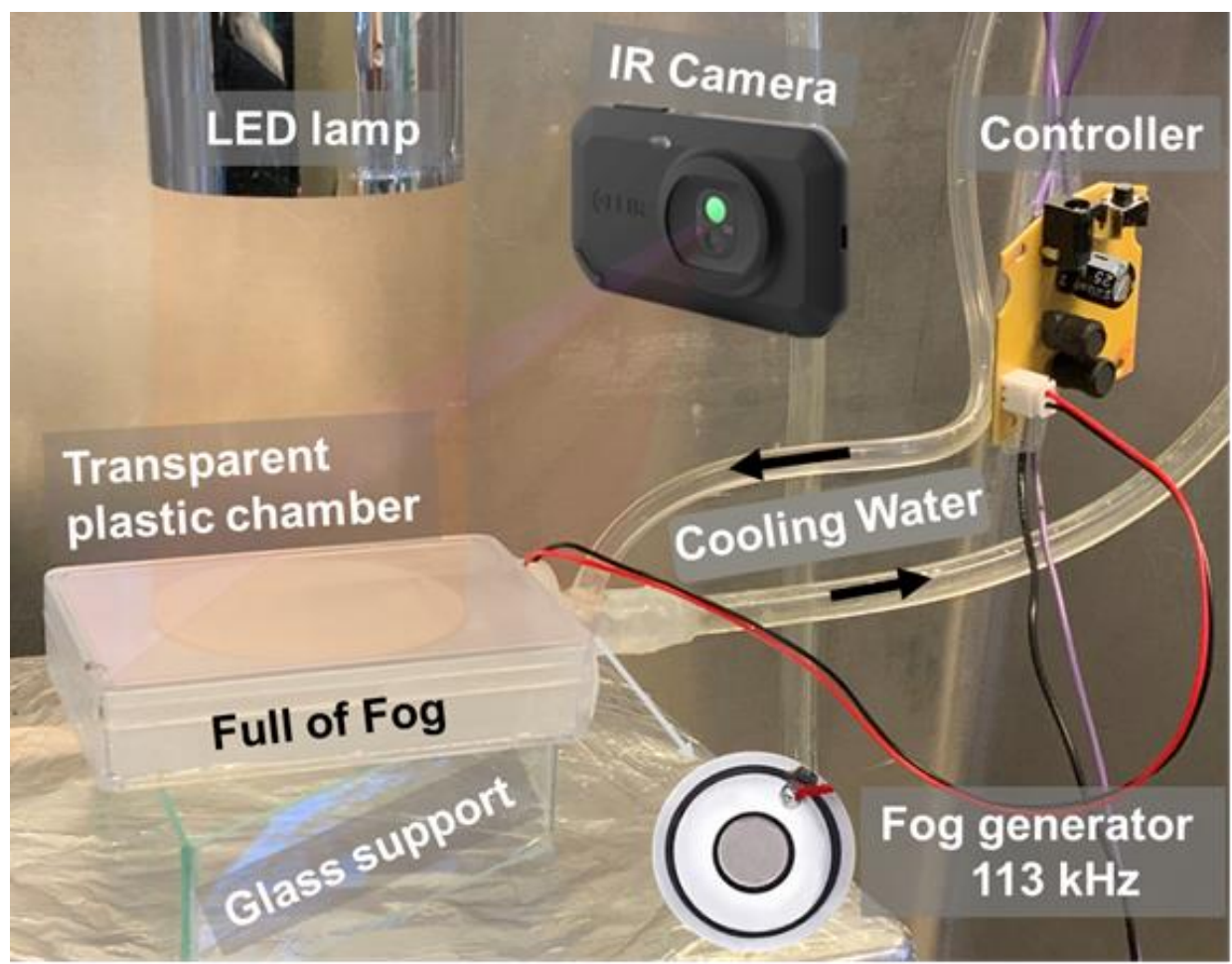

**Fig. S29. Fog experimental setup.** Fog generated by an ultrasonic vibrator is injected into a plastic container. The ultrasonic vibrator is cooled to avoid heating the fog.

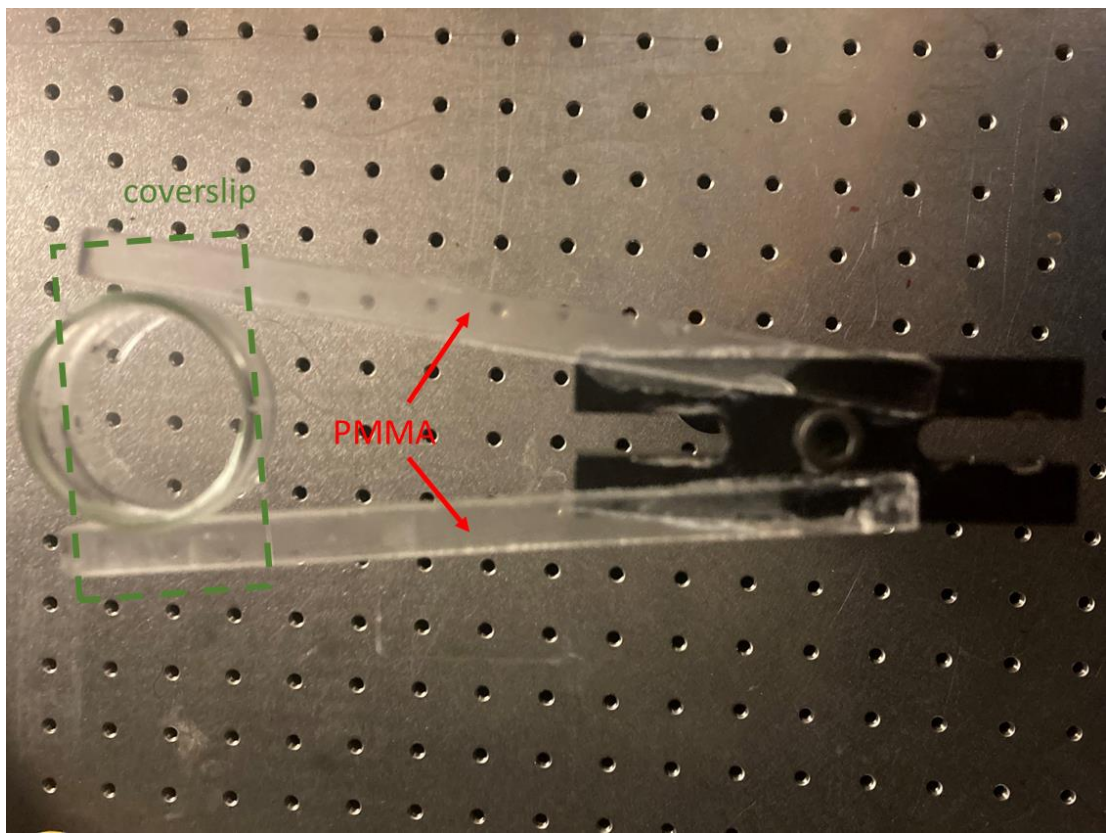

**Fig. S30. Set-up of suspended glass container.**

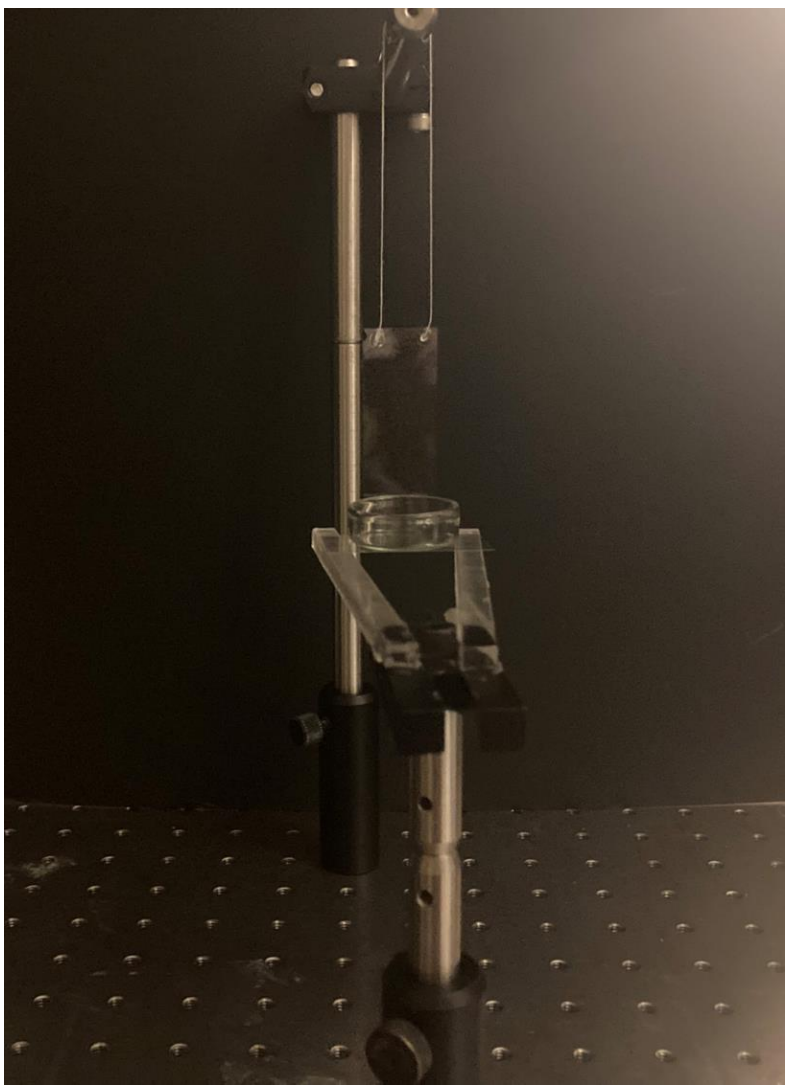

**Fig. S31. Set-up of vertical vapor phase temperature distribution.**

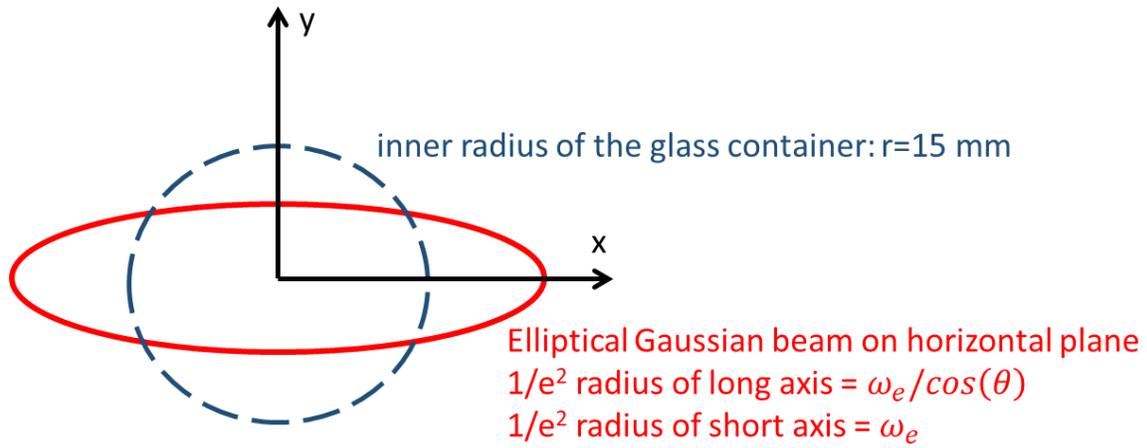

**Fig. S32. Geometry of expanded green laser and air-water interface on horizontal plane.** r is radius of air-water interface, $\omega_e$ is the beam radius of the expanded green laser (circular).

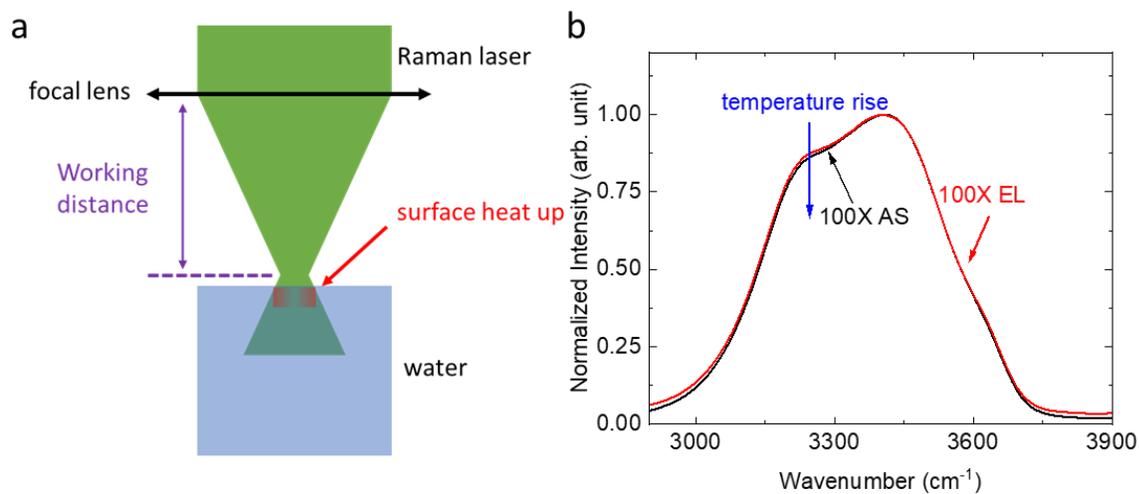

**Fig. S33. Raman thermometer measurement.** (a) Schematic of water surface heat up by photomolecular effect under Raman laser illumination; (b) Raman spectrum above water surface (100X EL and 100X AS). Working distance of 100X lens is 0.21 mm. The $1/e^2$ radius of green laser (532 nm) at beam waist is ~0.5 μm. Surface temperature of water is ~20.5 °C before laser turns on. The power of laser is 17 mW. Black curve is measured with the average distance of beam waist over water ~30 μm.

**Table S1. Power of expanded green laser (1.4 W) at air-water interface**

| $\theta$(°) | $P_{interface}$ (W) | Normalized intensity $(P_{interface}/S_{interface})$ (W/m$^2$) | Equivalent Intensity $(P_{interface}/S_{interface}/\cos(\theta))$ (W/m$^2$) |
|---|---|---|---|
| 0 | 1.3995 | 1981 | 1981 |
| 15 | 1.399 | 1980 | 2050 |
| 30 | 1.398 | 1979 | 2284 |
| 45 | 1.390 | 1967 | 2781 |
| 60 | 1.323 | 1873 | 3742 |
| 75 | 0.956 | 1353 | 5215 |

*The radius of air-water interface is 15 mm. The 1/e$^2$ radius of expanded laser is 7.5 mm.

**Table S2. Wavelength, 1/e² beam radius and maximum power of the pen lasers**

| Wavelength (nm) | Long axis 1/e² beam radius (mm) | Short axis 1/e² beam radius (mm) | Maximum power of pen laser (mW) |
|---|---|---|---|
| 450 | 1.6 | 0.5 | 3.5 |
| 520 | 2.3 | 0.85 | 4.5 |
| 635 | 1.9 | 0.64 | 2.5 |
| 850 | 1.9 | 0.75 | 3.4 |

**Table S3. Beam deflection/power of beam deflection measurement**

| Wavelength (nm) | Polarization | Angle (º) | Deflection/Power (μrad/mW) | Sample to Detector Distance (m) |
|---|---|---|---|---|
| 450 | TM | 15 | 2.7 | 0.1 |
| 450 | TM | 30 | 12.6 | 0.12 |
| 450 | TM | 45 | 49.1 | 0.18 |
| 450 | TM | 60 | 46.7 | 0.32 |
| 450 | TM | 75 | 49.6 | 0.4 |
| 520 | TM | 15 | 3.2 | 0.1 |
| 520 | TM | 30 | 10.9 | 0.12 |
| 520 | TM | 45 | 36.8 | 0.18 |
| 520 | TM | 60 | 41.2 | 0.32 |
| 520 | TM | 75 | 43.9 | 0.4 |
| 635 | TM | 15 | 3.1 | 0.1 |
| 635 | TM | 30 | 14.9 | 0.12 |
| 635 | TM | 45 | 55.9 | 0.18 |
| 635 | TM | 60 | 53.6 | 0.32 |
| 635 | TM | 75 | 50.5 | 0.4 |
| 850 | TM | 15 | 1.63 | 0.1 |
| 850 | TM | 30 | 5.8 | 0.12 |
| 850 | TM | 45 | 11.2 | 0.18 |
| 850 | TM | 60 | 17.3 | 0.32 |
| 850 | TM | 75 | 8.8 | 0.4 |

**Table S4. Beam deflection/power of beam deflection measurement**

| Wavelength (nm) | Polarization | Angle (º) | Deflection/Power (μrad/mW) | Sample to Detector Distance (m) |
|---|---|---|---|---|
| 450 | TE | 15 | 0.96 | 0.1 |
| 450 | TE | 30 | 3.31 | 0.12 |
| 450 | TE | 45 | 5.72 | 0.18 |
| 450 | TE | 60 | 4.57 | 0.32 |
| 450 | TE | 75 | 9.53 | 0.4 |
| 520 | TE | 15 | 0.8 | 0.1 |
| 520 | TE | 30 | 1.88 | 0.12 |
| 520 | TE | 45 | 4.71 | 0.18 |
| 520 | TE | 60 | 2.63 | 0.32 |
| 520 | TE | 75 | 7.31 | 0.4 |
| 635 | TE | 15 | 1.07 | 0.1 |
| 635 | TE | 30 | 3.1 | 0.12 |
| 635 | TE | 45 | 6.98 | 0.18 |
| 635 | TE | 60 | 10.3 | 0.32 |
| 635 | TE | 75 | 0 | 0.4 |
| 850 | TE | 15 | 3.75 | 0.1 |
| 850 | TE | 30 | 0.8 | 0.12 |
| 850 | TE | 45 | 3.68 | 0.18 |
| 850 | TE | 60 | 4.33 | 0.32 |
| 850 | TE | 75 | 0.8 | 0.4 |

**Table S5. Signal between 0f to 3f signal at air-water interface in the 2ω Raman measurement**

|  | 0f | 1f | 2f | 3f |
|---|---|---|---|---|
| Laser pump | ■ | ■ |  | ■ |
| Laser probe | ■ | ■ |  | ■ |
| Heating | ■ | ■ |  | ■ |
| Cluster | ■ | ■ |  | ■ |
| Raman (water) | ■ | ■ |  | ■ |
| Raman (cluster) | ■ | ■ | ■ | ■ |

**Table S6. Dependence and response of the measurements**

| Experiments \ Observations | polarization | wavelength | angle | molecular clusters | temperature response |
|---|---|---|---|---|---|
| surface temperature | ■ | | ■ | | ■ |
| vapor temperature | ■ | | | | ■ |
| surface beam deflection | ■ | | ■ | | |
| SS-TBD | ■ | ■ | ■ | | |
| vapor temperature (hot water) | ■ | | | | ■ |
| vapor transmission (hot water) | | | | ■ | |
| Raman thermometer | | | | | ■ |
| vapor phase Raman (hot water) | | | | ■ | |
| vapor phase Raman (cold water) | | | | ■ | |
| 2ω Raman | ■ | | | | |
| bending cantilever | ■ | | | | |
| evaporation | ■ | | | | |
| cloud experiment | | ■ | | | ■ |
| Raman thermometer | | | | | ■ |